\documentclass[reprint,aps,prd,nofootinbib,a4paper,10pt,twocolumn,superscriptaddress]{revtex4-2}
\usepackage{hyperref} 
\usepackage{changes}
\usepackage{graphicx,subfigure}
\usepackage[section]{placeins}
\usepackage[titletoc]{appendix}
\usepackage{xcolor}
\usepackage{dsfont}
\usepackage{bm}
\usepackage{mathtools} 
\usepackage{amsfonts}
\usepackage{enumerate}
\hypersetup{
     colorlinks   = true,
     citecolor    = cyan,
     linkcolor = pink
     }

\begin{document} 

\title{Testing inhomogeneous cosmography in our cosmic neighborhood using CosmicFlows-4}

\author{S. M. Koksbang} 
\email{koksbang@cp3.sdu.dk}
\affiliation{CP$^3$-Origins, University of Southern Denmark, Campusvej 55, DK-5230 Odense M, Denmark}

\begin{abstract} 
The convergence of the third order general cosmographic expansion of the luminosity distance is examined using several versions of a semi-realistic model of our local cosmic neighborhood, based on publicly available density and velocity fields from CosmicFlows-4. The study supports earlier findings that the general cosmographic expansion diverges at surprisingly low redshifts, often well before $z \sim 0.1$. By being based on a realistically placed observer within a data-informed cosmic environment, the results underscore that convergence must be a central concern when applying the general cosmographic expansion. By showing all-sky maps of kinematic parameters, the study also highlights the substantial information we lose when relying solely on standard FLRW-based cosmography.
\newline\indent
Poor convergence does not necessarily render the information extracted by fitting data to the general cosmographic expansion meaningless. Rather, it calls for caution in interpreting this information, particularly regarding the physical meaning of the fitting coefficients, the physical scales they probe and the implicit smoothing introduced by the fit.
\end{abstract}

\maketitle

\section{Introduction}
The $\Lambda$CDM model is the onset for interpreting observations in standard cosmology. However, observational data has become so ample and precise that fluctuations in observations across the sky can be studied for a variety of observables. An interesting example is the recent array of studies into the anisotropy of the Hubble diagram \cite{dipoleH1, dipoleH2, dipoleH3, dipoleH4, dipoleH5, dipoleH6, dipoleH7, dipoleH8, dipoleH9} (see e.g. \cite{dipoleH_summary} for an overview of the results of these studies and \cite{review} for a review). These studies illustrate the high precision of current cosmological observational data, demonstrating that the data has reached a level of detail that permits us to thoroughly study its anisotropy and inhomogeneity. Additionally, in recent years, observational challenges for the standard $\Lambda$CDM model have appeared. A striking but somewhat overlooked example is an apparent disagreement between the dipole anisotropy of the cosmic microwave background and other distant sources \cite{ani1, ani2, ani3} (see e.g. \cite{ani_original} for the background for these studies).
\newline\indent
Overall, one may argue that the time has come to go beyond the standard framework of employing Friedmann-Lemaitre-Robertson-Walker (FLRW) cosmology when interpreting observations. This is further supported by studies such as \cite{bulkflow} claiming strong tension between the local universe with the $\Lambda$CDM model. Such studies may indicate that the $\Lambda$CDM model and the corresponding standard tools for data analysis simply are not accurate enough to be used with modern astrophysical data, especially when considering low-redshift observations.
\newline\indent
Fittingly, there has been an increased activity among cosmologists in developing formalisms for taking anisotropy and inhomogeneity into account in data analyses. A prime example is the cosmographic expansion of the luminosity distance for a general spacetime presented in \cite{asta_cosmo}\footnote{Some of the results presented in \cite{asta_cosmo} were earlier derived in the PhD thesis \cite{thesis}.}, which has recently been augmented in the series \cite{augment1, augment2, augment3} and which builds on earlier work such as \cite{early1, early2, early3}. The formalism of \cite{asta_cosmo} has been used in studies based on numerical relativity \cite{ETdipole1, ETdipole2} and in N-body simulations based on the weak-field approximation \cite{gevolutiondipole}. The formalism has even been used in relation to studying real data in e.g. \cite{dipoleH3, dipoleH_summary}. However, already in \cite{ETdipole1, ETdipole2} it was noticed that the general cosmographic framework may have an obstacle in terms of a very low radius of convergence and/or a very slow convergence rate compared to the FLRW expansions (see e.g. \cite{cosmoFLRW1, cosmoFLRW2} regarding convergence of cosmographic expansions based on FLRW spacetimes). Indeed, good convergence at third order for redshifts up to $z \approx 0.1$ was only obtained in simulations with a smoothing scale as large as 200Mpc/h. The obstacle was further confirmed in \cite{asha} where the convergence of the general cosmographic expansion was studied in various Lemaitre-Tolman-Bondi (LTB) \cite{LTB1, LTB2, LTB3} models constructed to resemble our cosmic neighborhood. This study made it clear that the convergence of the underlying Taylor expansion of the general cosmographic expansion depends strongly on the local environment and that assessments stating that the expansion is suitable up to around $z = 0.1$ (as in e.g. \cite{augment3}) are simply too optimistic\footnote{Note added after publication: This assertion refers specifically to the convergence of the underlying Taylor expansion in a strict cosmographic sense. If, instead, one makes an empirical polynomial fit to data, this polynomial can of course easily remain an excellent approximation within the redshift range $z\in[0,0.1]$. While such a polynomial fit is not \emph{a priori} guaranteed to correspond to a genuine Taylor/cosmographic expansion, it is natural to expect that we may relate the fitted parameters to cosmographic parameters of a smoothed spacetime. This expectation provided the basis for \cite{augment3} and the point is discussed further here where the aim is to provide an initial investigation into the relation between polynomial-fitted values of cosmographic parameters and the underlying actual spacetime parameters of the spacetime smoothed on various scales, and how this relates to the convergence of the Taylor expansions.}.
\newline\indent
The studies listed above {\em indicate} that the general cosmographic expansion only converges at redshifts much lower than $z = 0.1$ unless structures are radically smoothed or expansion terms far beyond third order are included. Firmly establishing realistic limits of the convergence will, however, require detailed and realistic models of our cosmic neighborhood. The goal here is to make exactly such a model and place realistic limits on the convergence of the general cosmographic expansion of the luminosity distance. More precisely, the objective here is to study the convergence of the general cosmographic expansion devised in \cite{asta_cosmo}, using an array of versions of a model of our local universe based on real observations combined with a weak-field approximation. The first section below presents the constructed models of our local universe in detail. Following that, section \ref{sec:light} provides an overview of the formalism used to calculate the redshift-luminosity distance relation and cosmographic expansion for the model. Section \ref{sec:results} presents the results from a numerical investigation based on the models, while section \ref{sec:summary} provides a summary, discussion and conclusions.

\section{Model setup}\label{sec:model_setup}
This section serves to introduce the model that will be studied in the later sections. The goal is to construct a realistic model of the local universe so that the model can be used to realistically estimate the convergence of the general cosmographic expansion of the luminosity distance. Several versions of a model based on the overall same data will be considered and used in order to understand the significance of model details.
\newline\newline\noindent
\begin{figure*}
    \centering
    \includegraphics[width=\columnwidth]{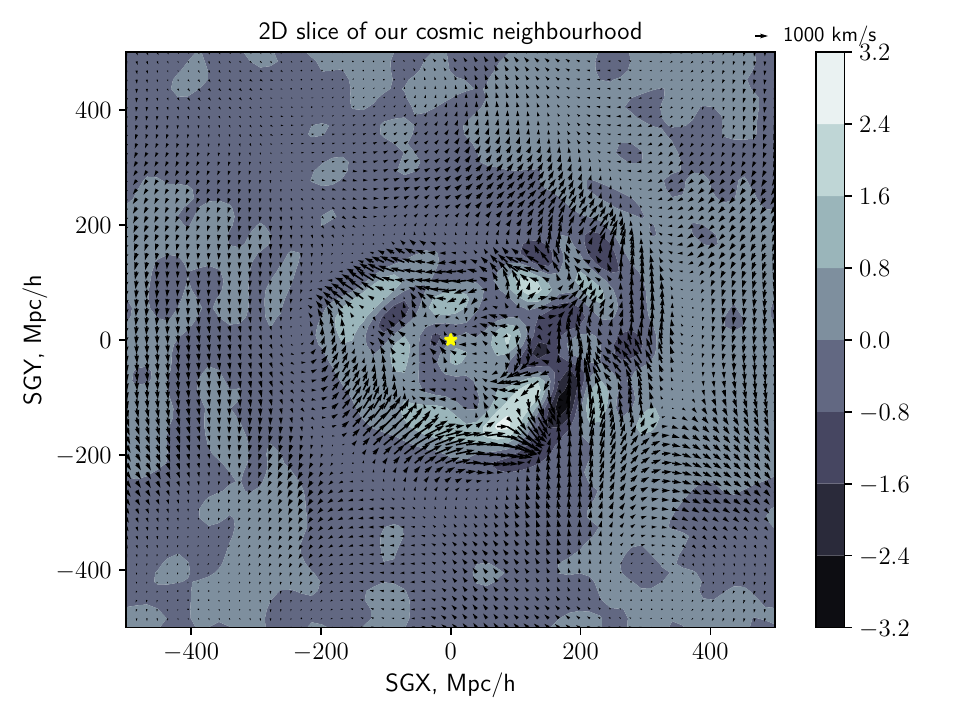}
        \includegraphics[width=\columnwidth]{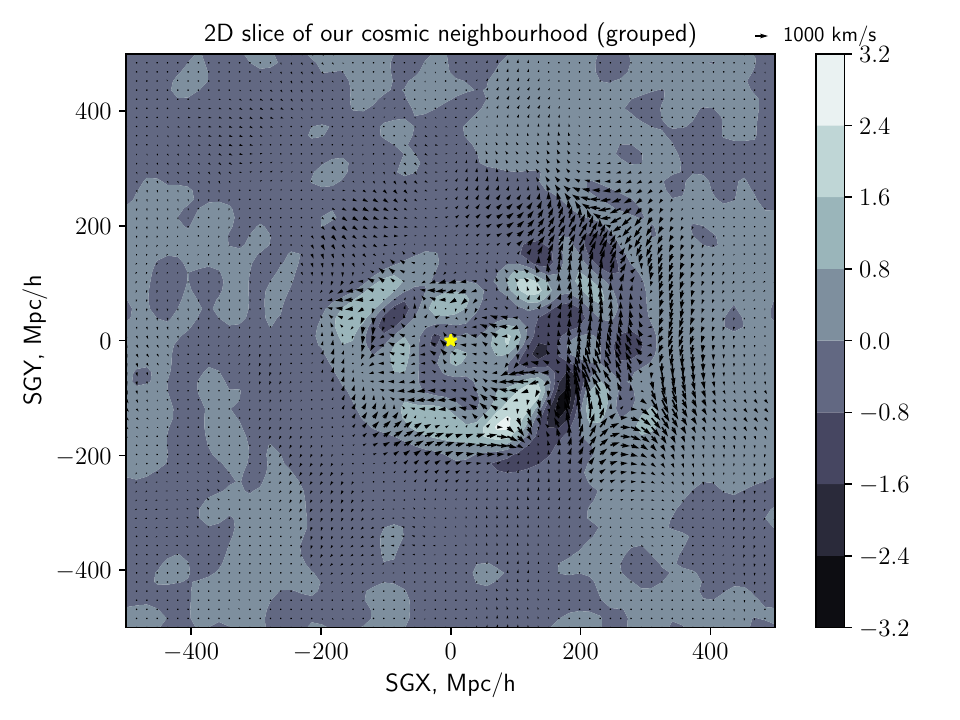}
    \caption{Two-dimensional slice of model universe corresponding to the middle plot in figure 2 of \cite{fields}. A star indicates our position at supergalactic coordinates (SGX, SGY, SGZ) = (0,0,0). The plot to the left shows the slice for ungrouped data while the plot to the right shows the fields for the grouped data. 
    }
    \label{fig:d2slice}
\end{figure*}
The realistic models are obtained by using the density and peculiar velocity fields presented in \cite{fields} based on the CosmicFlows-4 data \cite{CosmicFlows4}. The CosmicFlows-4 dataset contains distance and redshift measurements of nearly 56,000 galaxies in our local cosmic neighborhood. Corresponding peculiar velocities can be extracted from the data by assuming a background FLRW model (see e.g. \cite{basin} for a discussion of this approach). Density fields can then be estimated based on perturbative expressions. This was done in \cite{fields}, with the resulting density and velocity fields being made publicly available\footnote{The data can be found at https://projets.ip2i.in2p3.fr/cosmicflows/.}. Two versions of the density and velocity fields are available: a grouped and ungrouped version. In the ``ungrouped'' dataset, the fields are determined based on each individual galaxy. In the ``grouped'' dataset, the galaxies are bunched into $\sim 38,000$ groups from which average velocity and density fields are determined (see e.g. \cite{group} for details on the grouping procedure). To assess the robustness of the results presented in section \ref{sec:results}, models based on both these datasets will be considered. Figure \ref{fig:d2slice} shows the density contrast $\delta:=\rho/\rho_{\rm bg}-1$ and the peculiar velocity field for each model in a 2-dimensional slice going through $(\rm SGX,SGY,SGZ) = (0,0,0)$ (our position in supergalactic (SG) coordinates). As discussed in \cite{fields}, the data is based on the flat $\Lambda$CDM model as the background with $\Omega_{m,0} = 0.3$ and $H_0 = 74.6\rm km/s/Mpc$, with the Hubble constant value being equal to the one extracted from the CosmicFlows-4 data in \cite{CosmicFlows4}. This is the background model assumed here, but note that smaller variations of background parameters such as $H_0$ are not crucial for the results presented further below.
\newline\indent
Spacetimes based on these density and velocity fields will be constructed by introducing a weak-field approximation inspired by \cite{gevolution}. For this, it is assumed that our local spacetime can be approximated by the perturbed FLRW line element in Newtonian gauge (considering only scalar perturbations)
\begin{align}
    ds^2 = -c^2(1+2\Phi)dt^2 + a^2(1-2\Phi)(dx^2+dy^2+dz^2).
\end{align}
It was argued in \cite{greenwald} that a Newtonian N-body simulation can be mapped into a perturbed FLRW spacetime by applying the above line element and combining it with the 4-velocity $u^\mu \propto (1,v^i)$, where $v^i$ is the peculiar velocity field obtained from the simulation. The metric perturbation $\Phi$ is then obtained from the density field of the simulation through the Poisson equation, $\nabla^2\Phi = 4\pi Ga^2(\rho-\rho_{\rm bg})$. This method has e.g. been found to apply well to LTB models reproduced with Newtonian N-body simulations \cite{mapLTB1, mapLTB2}. A similar mapping will be used here, using the peculiar velocity field and density contrast from the CosmicFlows-4 data instead of simulation data. The spacetime will be considered using the weak field approximation, only keeping the leading order terms. Following table 1 of \cite{gevolution} we will thus neglect terms containing the potential except its second order spatial derivatives which are kept at linear order. The density contrast (proportional to the second derivative of the metric potential) and the peculiar velocity field and its derivatives are also kept at linear order.
\newline\indent
The velocity and density data of \cite{fields} is given on the interval $[-500,500]\rm Mpc/h$ on a 64-point grid in each direction. The gridded data must be interpolated to obtain a smooth field for the raytracing and computation of cosmographic parameters. Since the chosen interpolation scheme may affect the results, three different interpolations schemes will be used to interpolate between grid points, namely linear interpolation, cubic interpolation and Steffen interpolation (cubic interpolation that enforces monotonic behavior between interpolation/grid points). Each choice of interpolation scheme leads to a distinct model of our local cosmic neighborhood.
\newline\indent
The results also depend on the grid size. Although current public data is limited to a $64^3$-point grid, higher-resolution versions ($128^3$ and $256^3$ grids) exist as well. However, it is actually lower resolution that would lead to better convergence of the cosmographic expansion. The relation between the accuracy of the general cosmographic expansion and the resolution of simulation data was assessed in appendix A of \cite{ETdipole1} where it was e.g. gauged that the general cosmographic expansion would be accurate at percent level in the interval $0.04\lesssim z \lesssim 0.15$ only if the first at least {\em eight} terms of the expansion were used together with a smoothing scale of $\sim 200$Mpc/h. Better accuracy can be obtained with fewer terms if the smoothing is done on larger scales and/or smaller redshift intervals are considered. To study the impact of grid resolution we therefore also introduce smoothed versions of the CosmicFlows-4 data using simple down-sampling where grid values of the coarser grid are obtained by averaging over the original grid values covering the same spatial region.
\newline\newline
The resulting models of our local cosmic neighborhood are clearly only approximate. One may for instance notice that the density contrast becomes smaller than -1 in figure \ref{fig:d2slice}, indicating a negative density field which is obviously un-physical. A density contrast below -1 is also seen in Figure 2 of \cite{fields} and is presumably due to a breakdown of the linear approximation used to go from peculiar velocity field to density contrast. However, the negative density does not have any significant consequences for the results (viz. no striking effects occur when comparing results along light rays that go through regions with negative density compared to light rays propagating through positive density only).
\newline\indent
A further approximation comes into the model from the uncertainty of distance measures which for the CosmicFlows-4 data typically have an error of $\sim 15\%$. This error propagates into the density and peculiar velocity estimates as discussed in e.g. \cite{basin, CosmicFlows4}. These uncertainties will not be taken into account here since it is expected that a 15\% change in the fields will not make a significant difference regarding the overall conclusions.
\newline\indent
Lastly, it is worth noting that the model as described so far does not included considerations of time-dependence. Since the region probed by the CosmicFlows-4 data is fairly small and light will only be propagated to $z = 0.1$ in the following, the time evolution can largely be ignored and the supergalactic grid described above will be considered a purely spatial grid, although the small time evolution occurring along light rays is automatically taken (roughly) into account on this grid. In the following section, explicit time derivatives of the velocity field will, however, be required, but since these are expected to be subdominant (see e.g. \cite{gevolutiondipole}), we will use the approximation $v_{,t} = 0$. To include a time evolution of the density field we could use linear perturbation theory to estimate a time dependence for the density contrast as $\delta_{,t} = -\delta\Theta \approx -\partial_iv^i$ (see e.g. chapter 10 in \cite{god_bog}), where $\delta\Theta$ is the fluctuation in the local expansion rate and the second equality is valid to first order in the weak-field approximation used here. It has been verified along individual light rays that the results presented further below do not depend significantly on whether or not the density contrast is explicitly evolved according to this equation when ray tracing.
\newline\newline
It is not the claim that the resulting models constitute exact replicas of our local neighborhood. The models are merely approximations of our local universe, expected to be realistic and detailed enough to capture the dominant characteristics of the redshift-distance relation and corresponding cosmographic expansions realistically through ray tracing.

\section{Light propagation}\label{sec:light}
This section serves to introduce the formalism used for ray tracing through the model universes as well as specifying the utilized general cosmographic expansion.
\begin{figure*}
    \centering
    \includegraphics[width=\columnwidth]{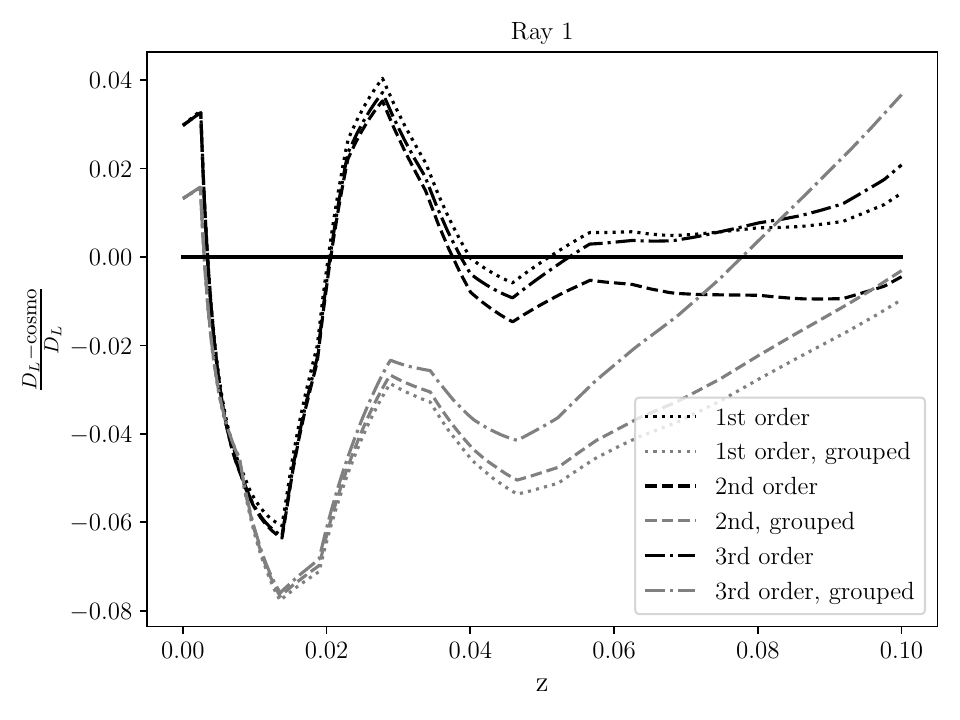}
        \includegraphics[width=\columnwidth]{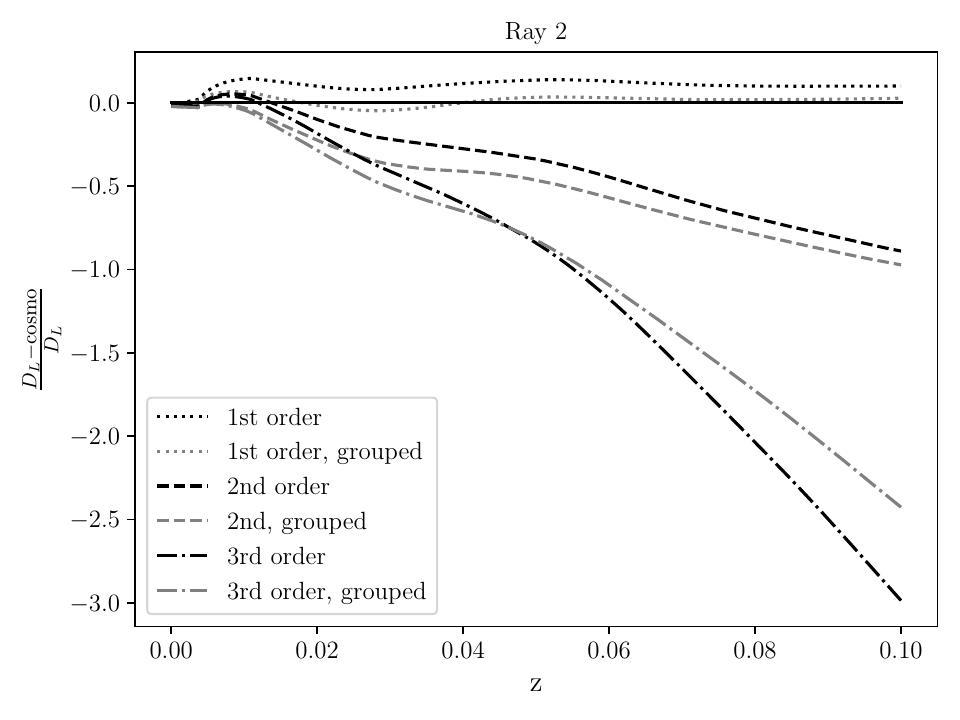}\\
            \includegraphics[width=\columnwidth]{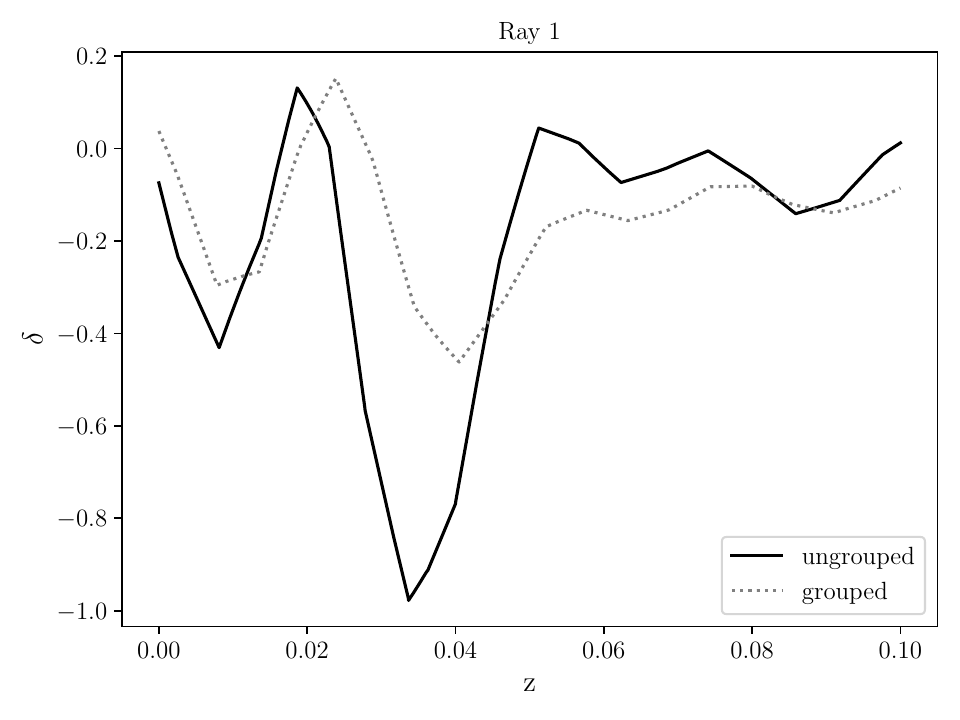}
        \includegraphics[width=\columnwidth]{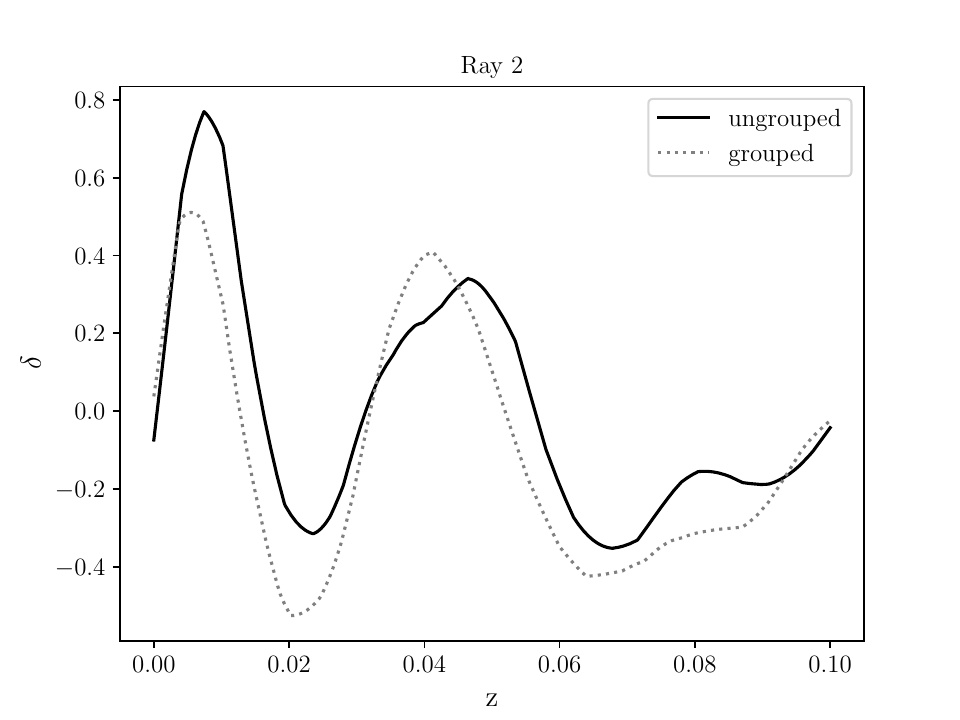}
    \caption{Density contrast and relative deviations between exact redshift-distance relation and cosmographic expansions (denoted as ``cosmo'') along two fiducial lines of sight in model based on linear interpolation and N = 64.
    }
    \label{fig:single}
\end{figure*}
\subsection{Cosmographic expansion}
Using a Taylor expansion in the redshift around $z = 0$, the luminosity distance can be written as
\begin{align}\label{eq:DL}
    D_L \approx D_L^{(1)}z + D_L^{(2)}z^2 + D_L^{(3)}z^3,
\end{align}
where the expansion coefficients $D_L^{(i)}$ depend on the chosen cosmological model. Following \cite{asta_cosmo} we will define generalized versions of the Hubble ($\mathcal{H}$), deceleration ($\mathcal{Q}$) and jerk ($\mathcal{J}$) parameters and a generalized spatial curvature term ($\mathcal{R}$). These are given by
\begin{align}\label{eq:coeff}
    \begin{split}
        \mathcal{H} &= \frac{1}{3}\Theta +e^\mu e^\nu \sigma_{\mu\nu}-\frac{e^\mu a_{\mu}}{c}\\
        \mathcal{Q} & = -1 - \frac{c^2}{E\mathcal{H}^2}\frac{d\mathcal{H}}{d\lambda}\\
        \mathcal{J} & = \frac{c^4}{E^2\mathcal{H}^3}\frac{d^2\mathcal{H}}{d\lambda^2}-4\mathcal{Q}- 3\\
                \mathcal{R} & = 1+\mathcal{Q} - \frac{c^4}{2E^2}\frac{k^\mu k^\nu R_{\mu\nu}}{\mathcal{H}^2},
    \end{split}
\end{align}
where $e^{\mu} = u^{\mu}/c - ck^\mu/(-u_{\alpha}k^{\alpha})$ is the spatial direction vector of the light ray as seen by an observer comoving with the dust, $\sigma_{\mu\nu}$ is the shear tensor and $\Theta$ the expansion rate of the matter. $E = -u^\alpha k_{\alpha} $ is the energy in the matter frame, and $a^\mu = u^\nu \nabla_\nu u^\mu$ is the 4-acceleration of the matter. Their explicit expressions in the models considered here are shown in appendix \ref{app:curl}. For the results shown in section \ref{sec:results}, the derivatives $d\mathcal{H}/d\lambda$ and $d^2\mathcal{H}/d\lambda^2$ were computed numerically using forward finite differences along the light rays. The former derivative is confirmed to be reliable by comparing it with its analytical expression while both derivatives are validated by comparing between finite differences of varying orders and different step sizes. 
\newline\indent
In \cite{asta_cosmo} it was shown that the coefficients of the cosmographic expansion for a general spacetime can be written as
\begin{align}
\begin{split}
    D_L^{(1)} &= \frac{c}{\mathcal{H_O}}\\
    D_L^{(2)} & = c\frac{1-\mathcal{Q_O}}{2\mathcal{H_O}}\\
    D_L^{(3)} & = c\frac{-1 + 3\mathcal{Q_O}^2+ \mathcal{Q_O - \mathcal{J_O}+\mathcal{R_O}}}{6\mathcal{H_O}},
\end{split}
\end{align}
assuming that the redshift is monotonous along the light ray, and where subscripted $\mathcal{O}$'s indicate evaluation at the observer.

\begin{table}[!htb]
\centering
\begin{tabular}{c c c c c c}
\hline\hline
Model & $D_L^{(1)}$ & $D_L^{(2)}$ & $D_L^{(3)}$ & $|D_L^{(1)}/D_L^{(2)}|$ & $|D_L^{(2)}/D_L^{(3)}|$\\ 
\hline
Ray 1, ungrouped & 4170  & 799  & -11228 & 5.2 & -0.0712  \\
Ray 1, grouped & $4314$ & $-280.8$ & -17164 & 15 & 0.0165 \\
Ray 2, ungrouped & 3899  & 42975  & 909204  & 0.0907 & 0.0473\\
Ray 2, grouped & 4239 & 43590 & 633489 & 0.0972 & 0.0688 \\
\hline
\end{tabular}
\caption{Expansion coefficients of two selected light rays in the grouped and ungrouped models. The fractions of neighboring coefficients are also shown.}
\label{table:coeff}
\end{table}
\subsection{Ray tracing}
To compute the redshift-distance relation using the models of our cosmic neighborhood introduced in section \ref{sec:model_setup}, we must solve the geodesic equations. Since the metric scalar perturbation and its first derivatives are neglected, the geodesic equations are simply those of the FLRW metric which we can write as
\begin{align}
\begin{split}
    \frac{dk^t}{d\lambda} &= -H(k^t)^2\\
    \frac{dk^i}{d\lambda} & = -2Hk^ik^t.
\end{split}
\end{align}
The luminosity distance, $D_L$, is obtained from the angular diameter distance, $D_A$, through the reciprocity relation. The angular diameter distance is computed as the square root of the determinant of the distortion matrix, $D$. The distortion matrix is obtained by solving the transport equation
\begin{align}
    \frac{d^2D}{d\lambda^2} = TD,
\end{align}
where $T$ is the tidal matrix with components based on $\mathbf{R}=-1/2R_{\mu\nu}k^{\mu}k^{\nu}$ and $\mathbf{F}=-1/2C_{\alpha\beta\gamma\mu}\left(\epsilon^*\right)^\alpha k^\beta \left(\epsilon^*\right)^\gamma k^\mu$ according to
\begin{align}
    T = \begin{pmatrix} \mathbf{R}- Re(\mathbf{F}) & Im(\mathbf{F}) \\ Im(\mathbf{F}) & \mathbf{R}+ Re(\mathbf{F})  \end{pmatrix}.
\end{align}
The transport equation is solved simultaneously with the geodesic equations and the parallel transport equation of the screen space basis vectors $E_1^\mu, E_2^\mu$ which are orthogonal to each other as well as $k^\mu$ and the observer velocity $u^\mu$\footnote{As discussed in \cite{sachs}, we may assume an arbitrary velocity field here and thus that the 4-velocity orthogonal to the screen space basis vectors is parallel transported along the light rays. Assuming this, we can obtain the basis vectors along the light rays by parallel transporting them along the null geodesics.}. The screen space basis vectors are combined into $\epsilon^\mu = E_1^\mu+iE_2^\mu$ when computing $\mathbf{F}$. However, the Weyl contribution to the angular diameter distance may be dropped since we are only interested in the dominant contributions to the distance fluctuations given by the gravitational and Doppler convergence, which are encapsulated already in $\mathbf{R}$ and the redshift $1+z = (u^\mu k_\mu)|_e/(u_\alpha k^\alpha)|_\mathcal{O}$. Indeed, at low redshift, we could even neglect the gravitational convergence and only consider the Doppler convergence (see e.g. \cite{bright_side, bonvin, mapLTB2} for discussions and demonstrations of this point). The results presented in the next section were obtained by considering both the Doppler and gravitational convergence, but it has been verified that the gravitational convergence is subdominant and that the results and conclusions presented below are unchanged if the gravitational convergence is neglected. For the model spacetime we have 
\begin{align}
    \mathbf{R} = -\frac{4\pi G\rho}{c^4} \left(u^\mu k_\mu\right)^2.
\end{align}

\begin{figure*}
    \centering
    \includegraphics[width=\columnwidth]{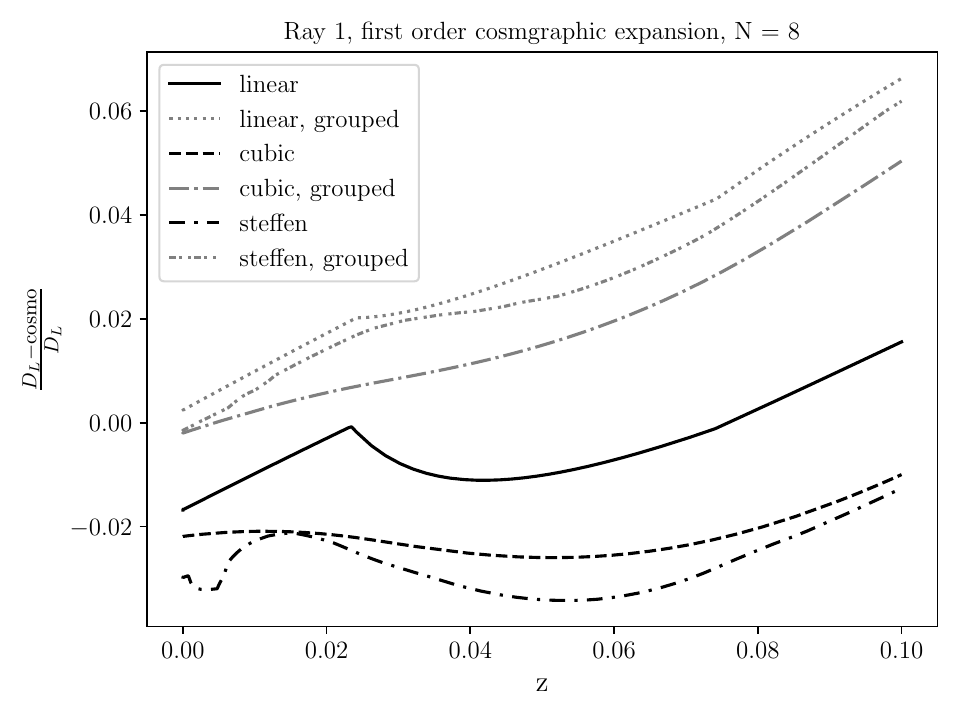} 
    \includegraphics[width=\columnwidth]{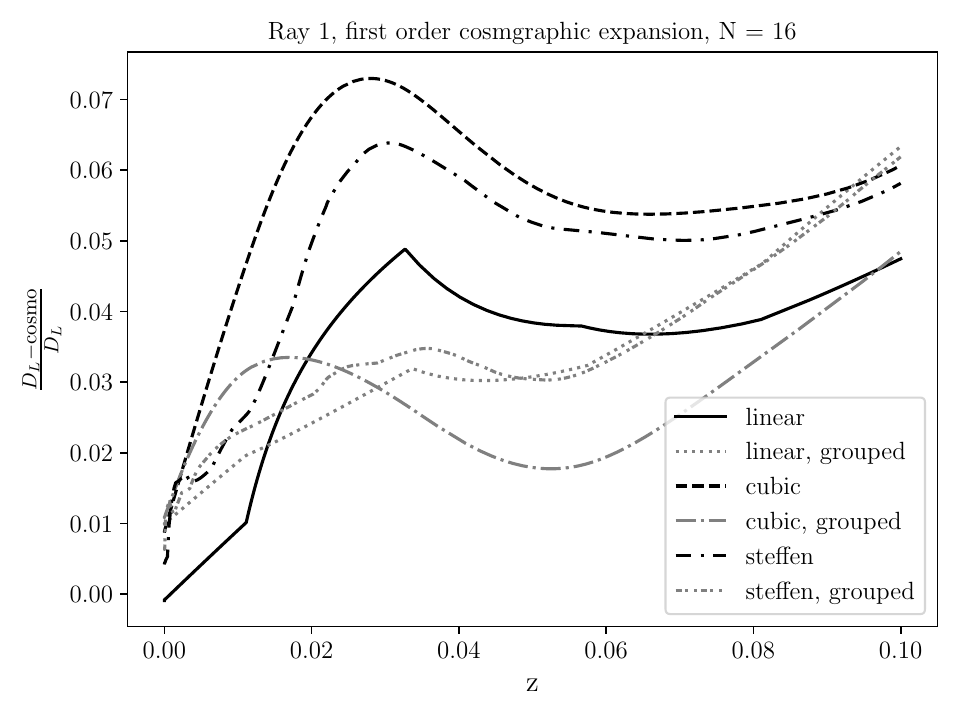}\\
    \includegraphics[width=\columnwidth]{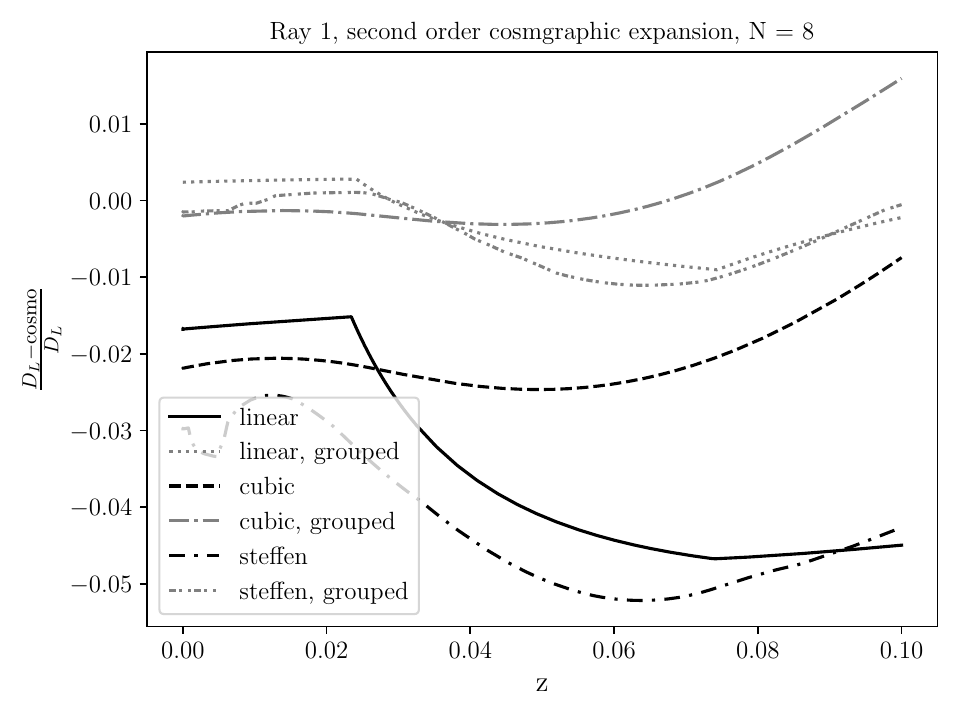}
    \includegraphics[width=\columnwidth]{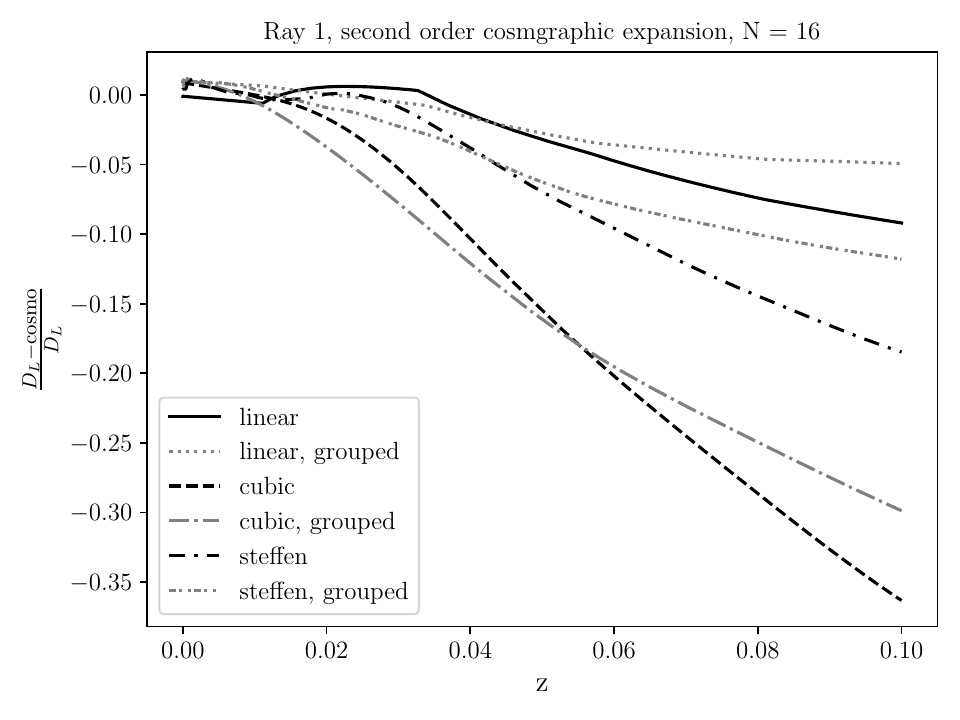}\\
       \includegraphics[width=\columnwidth]{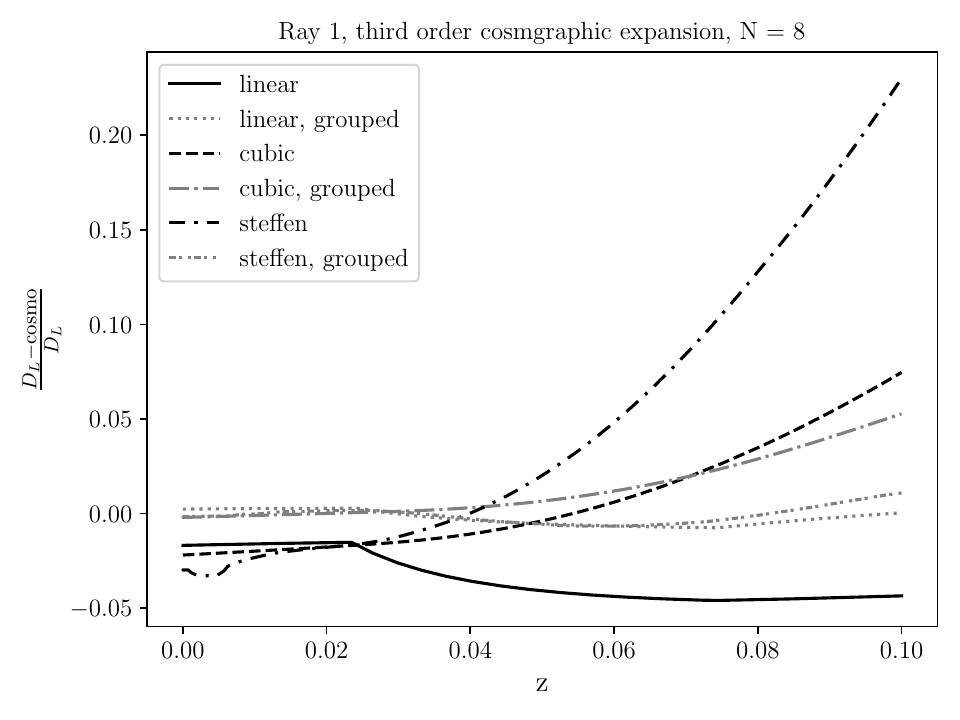}
        \includegraphics[width=\columnwidth]{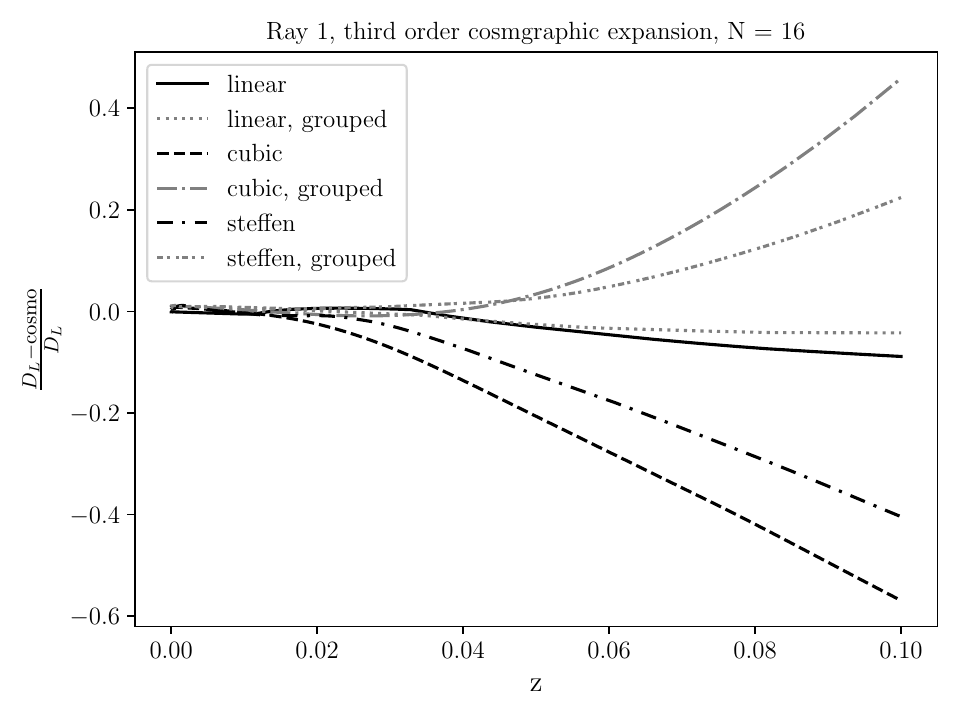}
    \caption{Relative deviation between the exact luminosity distance ($D_L$) and the cosmographic expansion (``cosmo'') at first, second and third order along ray 1 using an N = 8 and N = 16, and different interpolation schemes. The results are shown both using grouped (``grouped'') and ungrouped (not indicated) data.
    }
    \label{fig:ray1}
\end{figure*}

\begin{figure*}
    \centering
    \includegraphics[width=\columnwidth]{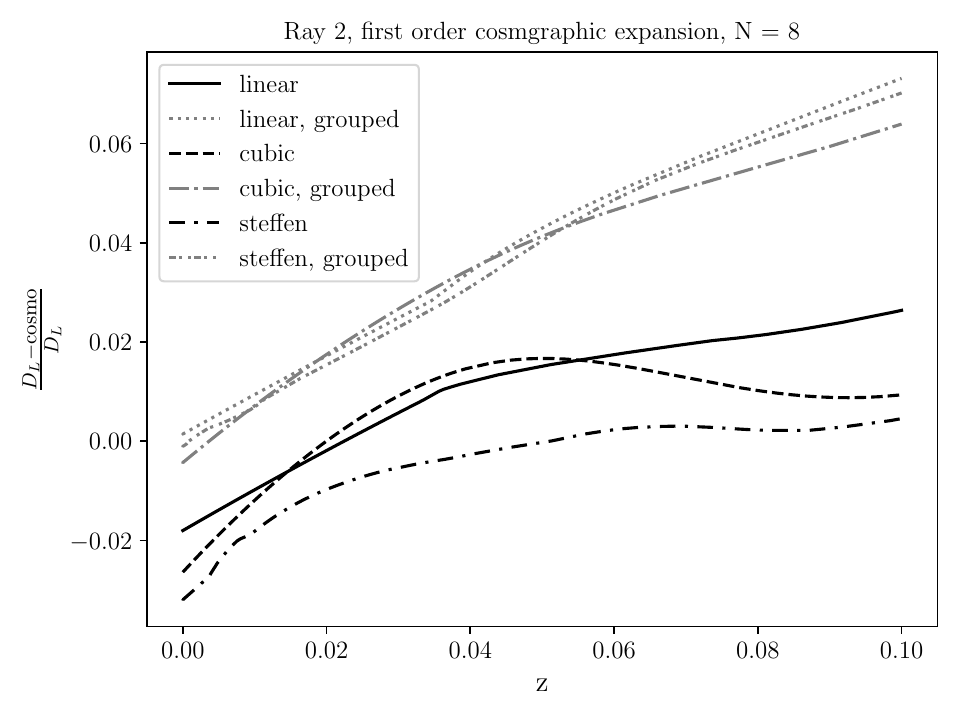} 
    \includegraphics[width=\columnwidth]{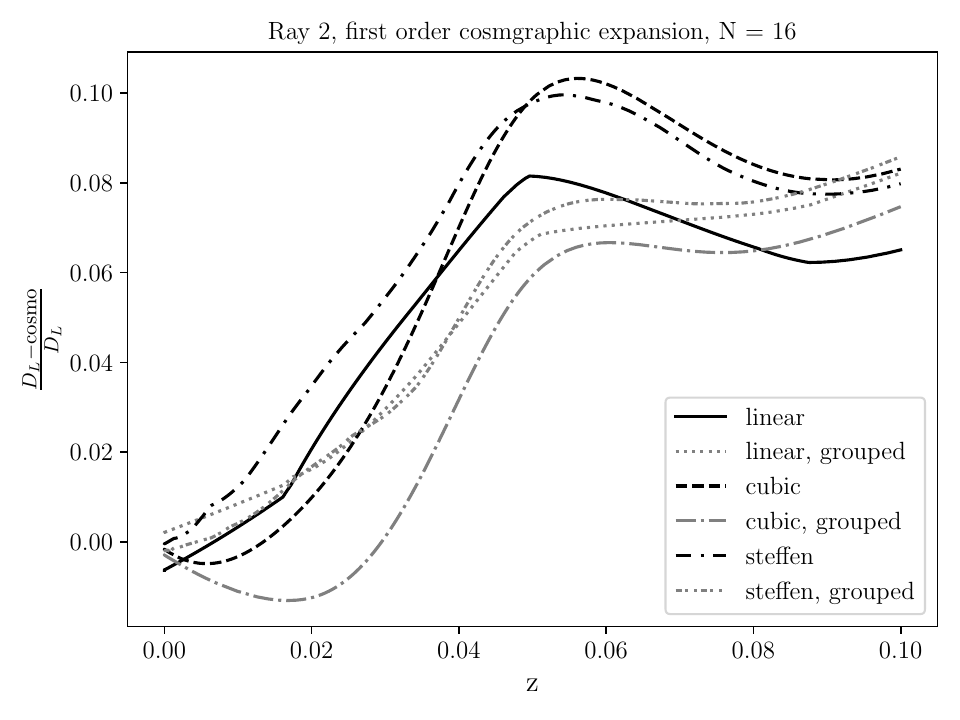}\\
    \includegraphics[width=\columnwidth]{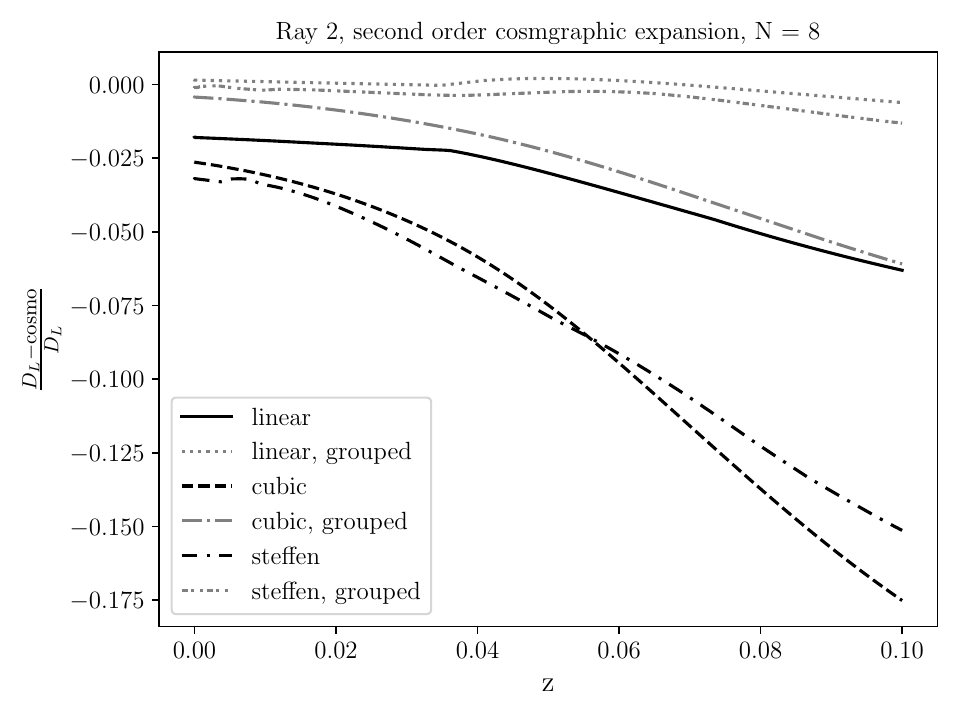}
    \includegraphics[width=\columnwidth]{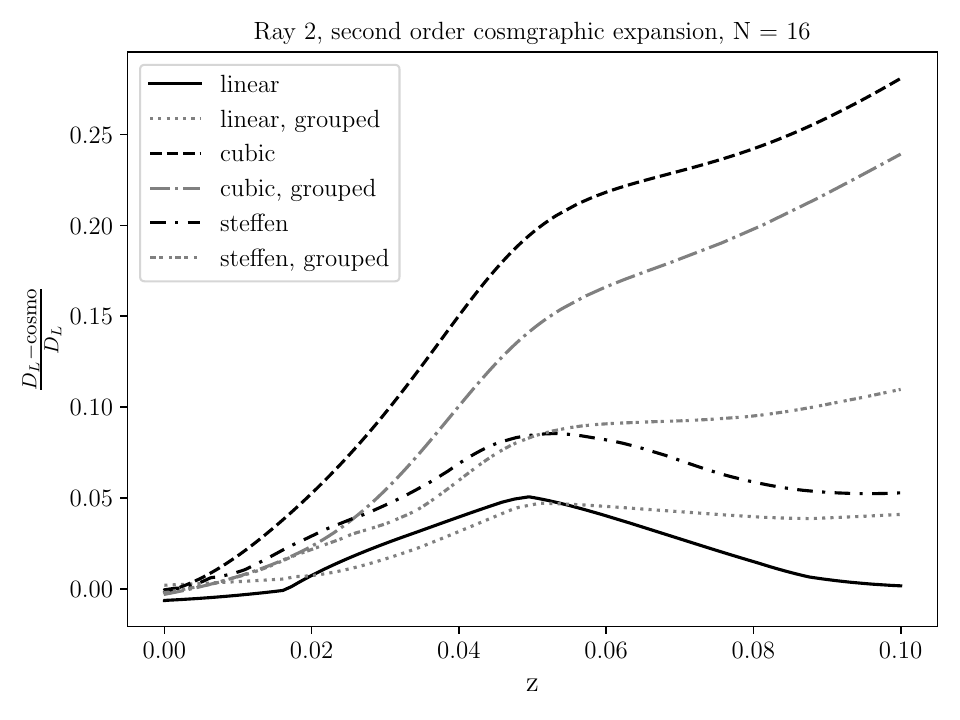}\\
       \includegraphics[width=\columnwidth]{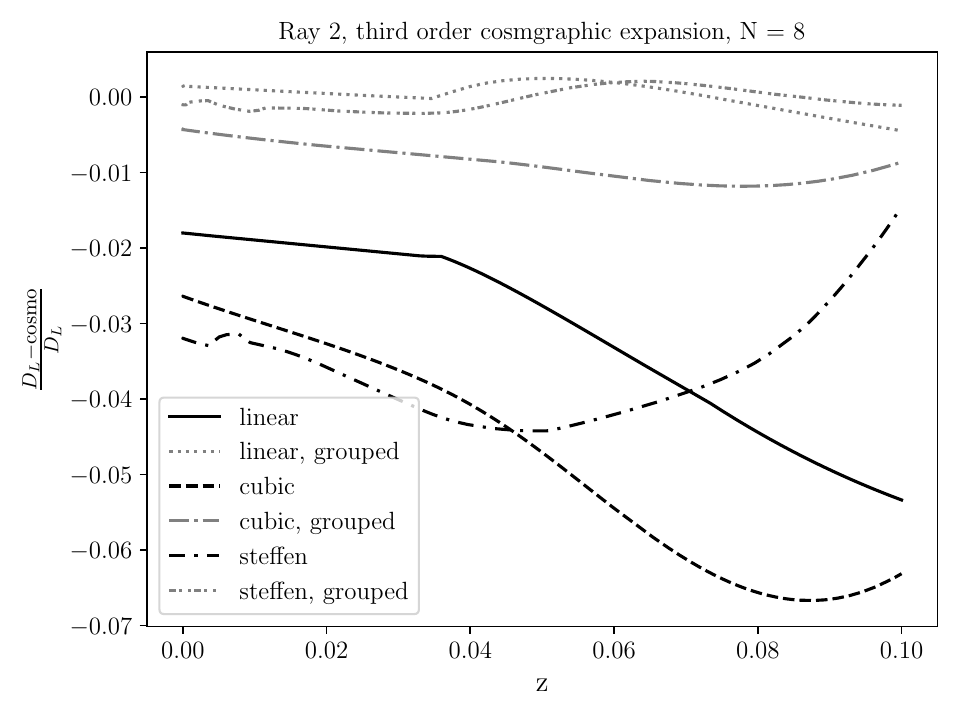}
        \includegraphics[width=\columnwidth]{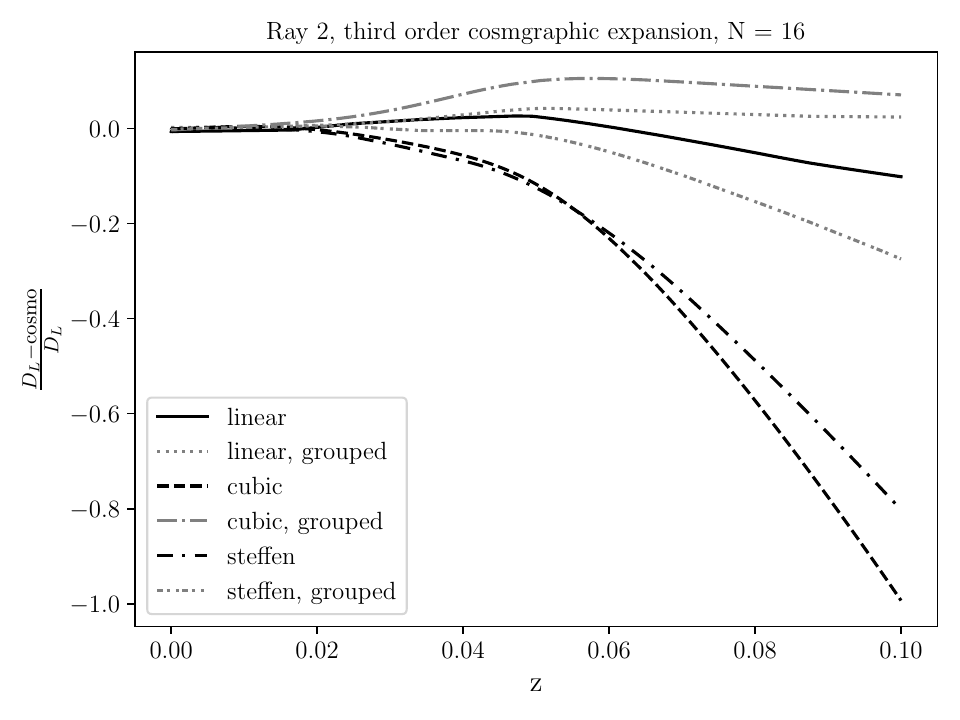}
    \caption{Relative deviation between the exact luminosity distance ($D_L$) and the cosmographic expansion (``cosmo'') at first, second and third order along ray 2 using grids with N = 8 and N = 16 and different interpolation schemes. Results are shown both using grouped (``grouped'') and ungrouped (not indicated) data.
    }
    \label{fig:ray2}
\end{figure*}
\begin{table*}[!h]
\centering
\begin{tabular}{c c c c c c c c c c}
\hline\hline
Model & $D_L^{(1)}$ & $D_L^{(2)}$& $D_L^{(3)}$& $\mathcal{H_O}$ (true) & $\mathcal{H_O}$ (fit) & $\mathcal{Q_O}$ (true) & $\mathcal{Q_O}$ (fit) & $\mathcal{J_O}$ & $\mathcal{R_O}$  \\ 
\hline
Ray 1 linear 64 & 4170 & 799 & -11228 & 71.90  & 70.96  & 0.617 & 1.675 & 18.1 & 1.168   \\
Ray 1 cubic 64 & 4165 & -53864 & -1230934  & 71.97 & 71.5 & 26.86 & 1.37 & 3991 & 27.4 \\
Ray 1 Steffen 64 & 4149 & -36981 & -5851997 & 72.25 & 71.45 & 18.82 & 1.40 & 9562 & 19.4 \\
\hline
Ray 1 linear 16 & 4043 & 5918 & -1358 & 74.15 & 70.57  & -1.93 & 1.842 & 8.87 & -1.400 \\
Ray 1 cubic 16 & 4000 & 18055 & 87562 & 74.95 & 68.5  & -8.028 & 2.780  & 45.49 & -7.46 \\
Ray 1 Steffen 16 & 3997 & 10307 & 93048 & 75.00 & 70.1  & -4.157 & 1.89 & -96.59 & -3.61 \\
\hline
Ray 1 linear 8 & 4205 & 2585 & -654 & 71.3 & 71.1  & -0.230 & 2.20 & 0.0486 & 0.1865 \\
Ray 1 cubic 8 & 4285 & -109 & -34812 & 69.97 & 70.8  & 1.05 & 2.17 & 53.41 & 1.30 \\
Ray 1 Steffen 8 & 4315 & 1280 & -116171 & 69.48 & 70.17  & 0.407 & 2.79  & 162.2 & 0.742 \\
\hline
Ray 1 linear 4 & 3985 & 3063 & -955&  75.2 &  74.99  & -0.537   &  -0.498  &  0.822  & 0.056   \\
Ray 1 cubic 4 & 3974 & 3061 & -2.30 &  75.4 &  75.11 &  -0.541& -0.557 & -0.580 & 0.08044  \\ 
Ray 1 Steffen 4 & 3975 & 3042 & 1613 & 75.43 & 75.28 & -0.531 & -0.708 & -3.04 & 0.0748 \\
\hline\hline
Ray 1 gp linear 64 & 4314 & -281 & -17164 & 69.49  & 72.10  & 1.13 & 2.239  & 29.4 & 1.592  \\
Ray 1 gp cubic 64 & 4347 & -33641 & -3487493  & 68.97 & 72.4 & 16.48 & 2.06 & 5661  &  16.9\\
Ray 1 gp Steffen 64 & 4399  & -36691 & -1416616 & 68.15 & 72.48 & 17.68 & 2.03 & 20227 & 18.1 \\
\hline
Ray 1 gp linear 16 & 4004 & 48112 & -2932 & 74.88 & 72.4 & -1.40 & 1.81 & 7.03 & -0.8677  \\
Ray 1 gp cubic 16 & 4078 & 14886 & -324610 & 73.51 & 69.9 & -6.3 & 3.43 & 583.6 & -5.76 \\
Ray 1 gp Steffen 16 & 4010 & 7686 & -146157 &  74.76 & 71.7 & -2.83 & 2.22 & 236.7 & -2.29 \\
\hline
Ray 1 gp linear 8 & 4001 & 2933 & -1134 & 74.9 & 73.7 & -0.466 & 1.01 & 0.892 & 0.005200 \\
Ray 1 gp cubic 8 & 4044 & 1465 & -15687 & 74.12 & 73.6 & 0.276 & 1.30 & 23.28 & 0.609 \\
Ray 1 gp Steffen 8 & 4014 & 2670 & -4901 & 74.68 & 73.2 & -0.33 & 1.62 & 6.42 & 0.0940 \\
\hline
Ray 1 gp linear 4 & 3986 & 3085 & -950 & 75.2 & 74.9 & -0.548 & -0.516 & 0.827 & 0.0449\\
Ray 1 gp cubic 4 & 3963 & 3034 & 1458 & 75.65 & 75.2 & -0.531 & -0.571 & -2.65 & 0.0912\\
Ray 1 gp Steffen 4 & 3967 & 3036 & 1714 & 75.58 &  75.20 & -0.531 & -0.656 & -3.20 & 0.0764\\

\hline\hline
Ray 2 linear 64 & 3899 &  42975 & 909204 & 76.88  & 70.63  & -21.04  & -2.29  & -113 & -20.4 \\
Ray 2 cubic 64 & 3888 & 125636 & 8360875 & 77.10 & 68.98 &-63.62 & -1.21  & -886 & -63.02 \\
Ray 2 Steffen 64 & 3819 & 127333 & 12965992 & 78.5 & 69.09& -65.68 & -1.28 & -7560 & -65.1 \\
\hline
Ray 2 linear 16 & 4036 & 2733 & 44512 & 74.28 & 74.94 & -0.354 & -4.45  & -66.97 & 0.174 \\
Ray 2 cubic 16 & 3932 & -8480 & 54367 & 76.24 & 80.92 & 5.31 & -9.05 & -739 & 5.89   \\
Ray 2 Steffen 16 & 3965 & 1159 & 368439 & 75.6 & 76.16 & 0.415 & -5.67 & -556.6 & 0.972 \\
\hline
Ray 2 linear 8 & 4231 & 3888 & -4215 & 70.85 & 71.57  & -0.838 & -0.247 & 5.81 & -0.429 \\
Ray 2 cubic 8 & 4333 & 8071  & -48996 & 69.18 & 69.56  & -2.73 & -0.342   & 83.9 & -2.49 \\
Ray 2 Steffen 8 & 4332 & 6784 & -59537 & 69.2 & 70.05 & -2.13 & -0.00400  & 91.2  & -1.80 \\
\hline
Ray 2 linear 4 & 4001 & 3288 & -1200 & 74.93 & 74.77 & -0.644 & -0.818 & -1.35 & -0.0527 \\ 
Ray 2 cubic 4 & 3995 & 3425 & -1750 & 75.03 & 74.58 & -0.714 & -0.666   & 2.35 & -0.0979 \\
Ray 2 Steffen 4 & 3997 & 3486 & -901 & 75.00 & 74.57& -0.744 & -0.693 & 1.13 & -0.143 \\
\hline\hline
Ray 2 gp linear 64 & 4239 & 43590 & 633489 & 70.72 & 74.38  & -19.57  & -2.90 & 213 & -19.1  \\
Ray 2 gp cubic 64 & 4288 & 131781 & 4784052 & 69.92 & 73.08 & -60.03 & -2.12 & 4154 & -60.02 \\
Ray 2 gp Steffen 64 &  4290 &  140084 & 5741691  & 69.9 & 73.23  & -63.87 & -2.23 & 4249  & -63.9  \\
\hline
Ray 2 gp linear 16 & 3979 & 1782 & 7292 & 75.35 & 76.67 & 0.104 & -3.53 & -11.2 & 0.646 \\
Ray 2 gp cubic 16 & 3994 & -7100 & 72659 & 75.06 & 80.35 & 4.56 & -6.16 & -38.2 & 5.11 \\
Ray 2 gp Steffen 16 & 3957 & -1038 & 166334 & 75.8 & 77.64 & 1.52 & -4.25 & -243 & 2.08 \\
\hline
Ray 2 gp linear 8 & 4030 & 3447 & -733 & 74.40 & 74.68 & -0.733 & -1.21 & 0.650 & -0.247 \\
Ray 2 gp cubic 8 & 4085 & 5448 & -22826 & 73.30 & 73.32  & -1.7 & -0.97  &  37.9  & -1.35 \\
Ray 2 gp Steffen 8 & 4042 & 3624 & -5241 & 74.2 & 74.83  & -0.793 & -1.45 & 7.50 & -0.377 \\
\hline
Ray 2 gp linear 4 &  4000 & 3221 & -1059 & 74.94 &74.67 & -0.611 & -0.613  & 1.08 & -0.0205   \\
Ray2 gp cubic 4 & 3985 & 3367 & -225 &  75.23  & 74.72   & -0.69  &  -0.661   & 0.00733  & -0.0709 \\
Ray 2 gp Steffen 4 & 3985 & 3354 & 11.90 & 75.22 &74.71 & -0.683 & -0.637 & -0.381 & -0.0795\\

\hline
\end{tabular}
\caption{Expansion coefficients for ray 1 and ray 2 as well $\mathcal{H_O}, \mathcal{Q_O}, \mathcal{J_O}$ and $\mathcal{R_O}$ for all considered models, i.e. models based on grouped (gp) and ungrouped data, using both linear, cubic and Steffen interpolation with N = 64, 16, 8 and 4. Values of $\mathcal{H_O}$ and $\mathcal{Q_O}$ obtained from polynomial fitting along the redshift-distance relation are also shown. These are indicated as ``fit'' while $\mathcal{H_O}$ and $\mathcal{Q_O}$ obtained by direct computation along the light rays are denoted as ``true''. $\mathcal{H_O}$ is given in km/s/Mpc.}
\label{table:fit}
\end{table*}

\section{Numerical results}\label{sec:results}
This section serves to present results from propagating light rays through the different versions of the model of our cosmic neighborhood introduced in section \ref{sec:model_setup}.
\newline\newline
We will begin by studying the more naive models based on linear interpolation on the original grid with $64^3$ grid points, using both the grouped and ungrouped data to see how grouping affects results. For this, we will consider two selected light rays (``ray 1'' and ``ray 2''), each highlighting important points. Figure \ref{fig:single} shows the cosmographic expansion relative to the exact redshift-distance relation along the two fiducial light rays in these two models. The plots to the left show the results from a light ray along which the deviation between the exact and cosmographic results remain fairly small for the entire studied redshift interval. There are noticeable deviations between the exact and cosmographic expansions but they are only of a few percent and do not grow significantly along the light ray, until at $z\approx 0.05$. These deviations are therefore expected to largely represent lack of precision of the model, i.e. imprecision due to computing derivatives using finite-difference formulas, linear interpolations on the grid, and neglecting higher-order terms. The plots to the right in figure \ref{fig:single} show a very different behavior. In these plots, the deviation between the cosmographic expansion and the exact redshift-distance relation grows strongly with the redshift. Along this line of sight it is clear that the cosmographic expansion breaks down long before $z = 0.1$ is reached. The figure to the left might similarly be revealing the beginning of divergence of the cosmographic expansion for $z\geq0.05$, where the deviation between the exact redshift-distance relation and the cosmographic expansions seem to begin to grow. The results are very similar for the grouped versus ungrouped model.
\newline\indent
Figure \ref{fig:single} also shows the density contrasts along the corresponding light rays. From this, it is clear that the density contrast has a much steeper gradient near the observer along the light ray where the cosmographic expansion most clearly seems to break down. It is not surprising that such strong gradients lead to a significantly reduced radius of convergence or slow convergence rate of a Taylor expansion. Table \ref{table:coeff} shows the expansion coefficients for the four light rays. The table also shows the fractions of the coefficients. These are included in the table because the radius of convergence, $R$, can be computed as $R = \lim\limits_{n \to \infty}|a_n/a_{n+1}|$, where $a_n$ are the coefficients of the expansion.
\newline\indent
Comparison of table \ref{table:coeff} and figure \ref{fig:single} reveals that the diverging cosmographic expansion appears along the line of sight where $D_L^{(3)}$ is about one order of magnitude larger than $D_L^{(2)}$ which again is one order of magnitude larger than $D_L^{(1)}$. In this case, the third order of the Taylor expansion begins to dominate over the first order expression when $z\approx \sqrt{D_L^{(1)}/D_L^{(3)}}\approx 0.065$ (using values for the ungrouped model). This value is clearly too optimistic when compared with figure \ref{fig:single} where the divergence appears to begin already at $z\approx 0.0075$. We can understand this simply as a reminder that the true radius of convergence cannot be estimated without including more expansion coefficients. This is also indicated by the coefficient fractions in table \ref{table:coeff}; to estimate the radius of convergence, the fractions would first need to converge towards a specific value which clearly has not happened in table \ref{table:coeff}.
\newline\indent
The same caveat remains for light ray 1 where we also cannot reliably estimate the radius of convergence. One may nonetheless note that for this line of sight, the third order term is closer to the same order of magnitude as the first order term, and the third order term begins to dominate over the first order term when $z\approx \sqrt{D_L^{(1)}/D_L^{(3)}}\approx 0.6$ (again using data for the ungrouped model). This is an order of magnitude larger than the estimate for light ray 2 which is consistent with the fact that the divergence happens at lower redshift along light ray 2 than along light ray 1.
\newline\indent
Since we cannot make a mathematically precise estimate of the radius of convergence, it is not possible to determine whether the problem actually is a very small radius of convergence or if it is merely a very slow rate of convergence. In the latter case, a better agreement between the cosmographic expansion and the exact redshift-distance relation would be achieved simply by adding more terms to the cosmographic expansion. Either way, it is important to understand to what extent the cosmographic expansion is affected by the chosen interpolation method and to what extent divergence of the cosmographic expansion can be remedied by using a coarser grid. This is therefore studied in the next subsection.

\subsection{Significance of grid resolution and interpolation scheme}
The same two light rays\footnote{Light rays are here consider ``the same'' if they have the same initial conditions, i.e. if they correspond to the same observer position and the same line of sight.}, ray 1 and ray 2, have been studied in the models based on linear, cubic and Steffen interpolation and with grids down-sampled to $16^3$, $8^3$ and $4^3$ grid points (hereafter referred to as N = 16, N = 8 and N = 4 grids, respectively). The former grid resolution corresponds to grid cells/a smoothing length of 62.5Mpc/h, N = 8 grid yields a smoothing length of 125Mpc/h and N = 4 grid yields a smoothing length of 250Mpc/h.
\newline\newline
Figures \ref{fig:ray1} and \ref{fig:ray2} compare the accuracy of the cosmographic expansions when using the different interpolation schemes with the N = 16 and N = 8 grids along the two light rays. Although the plots from the different models do not coincide exactly, the difference between them is not too striking. In all cases, the cosmographic expansions become visibly more accurate when the grid is coarsened. For the N = 16 grid, the third order cosmographic expansions now reproduce the exact redshift-distance relation to around percent precision up to $z\approx 0.02$. At higher redshifts, the cosmographic expansions begin to deviate significantly from the exact relation. For the N = 8 grid, the third order cosmographic expansions agree with the exact redshift-distance relation within 10 percent up to redshift $z\approx 0.1$ (the entire studied redshift interval), except for the ungrouped data with steffen interpolation along ray 1. The accuracy of the cosmographic expansion is around 1\% for several versions of the ray in the entire redshift interval. Along ray 2 there is a general trend of the cosmographic expansion to become better at higher order when the N = 8 grid is used, but this trend is {\em not} seen along ray 1. This indicates that even when using a smoothing length of 125Mpc/h, the cosmographic expansion is not reliable in giving accurate reproductions of the true redshift-distance relation.
\newline\indent
For the interested reader, the densities along the light rays using the different interpolations schemes and the grids N = 8 and N= 16 are shown in appendix \ref{app:density}.
\newline\newline
The cosmographic expansions not only deviate from the exact redshift-distance relation but also to some extent from each other, implying that their respective cosmographic parameters are not the same. This is confirmed in table \ref{table:fit} which shows the cosmographic coefficients and the parameters $\mathcal{H_O}, \mathcal{Q_O}, \mathcal{J_O}$ and $\mathcal{R_O}$. The table reveals that the cosmographic coefficients $D_L^{(2)}$ and $D_L^{(3)}$ in some cases vary significantly when changing the interpolation scheme or grid resolution and can even change sign. The value of $D_L^{(1)}$ is much more ``stable'', i.e. it changes much less when changing interpolation scheme and/or grid resolution. This is because $D_L^{(2)}$ and $D_L^{(3)}$ depend on higher order derivatives and these change significantly when the grid resolution and/or interpolation scheme is changed.
\newline\indent
It is expected that the cosmographic coefficients change at least when the grid resolution is, since the spacetime along the individual light rays depends strongly on the grid coarseness. The more important result to highlight is therefore that the value of $\mathcal{H_O}$ is fairly stable when comparing results from models with the same grid resolution but different interpolation schemes. The same is not true for $\mathcal{Q_O}$ which can even change sign when changing interpolation scheme. When using lower grid resolution this is mitigated but even for $N = 4$ the values of $\mathcal{Q_O}$ obtained using the different interpolation schemes vary by the order of 10\%.

\subsubsection{A comment on polynomial fitting}
The fundamental question this manuscript seeks to address is whether the Taylor expansion converges in realistic, anisotropic models based on observational data. This question can only be meaningfully answered by comparing the actual Taylor series (using the true cosmographic coefficients) to the exact redshift-distance relation. However, when dealing with real-world data these true coefficients are unknown. Indeed, the goal of a cosmographic analysis of real-world data would be to constrain the cosmographic parameters. In practice this would occur by fitting the data to a third order polynomial of the form $p(z):=c_1z + c_2z^2+c_3z^3$. 
Table \ref{table:fit} therefore also includes coefficients from fitting a third order polynomial of the form $p(z)$ to the individual redshift-distance relations. Since polynomials typically provide good fits to smooth data, such a fit will clearly be very accurate. However, since the third order Taylor expansions do not accurately reproduce the exact redshift-distance relation, the fitted coefficients $c_1$, $c_2$, and $c_3$ will not generally approximate the corresponding Taylor expansion coefficients $D_L^{(1)}$, $D_L^{(2)}$, and $D_L^{(3)}$. This mismatch implies that the values of $\mathcal{H_O}, \mathcal{Q_O}, \mathcal{J_O},$ and $\mathcal{R_O}$ derived from a polynomial fit will not, in general, reflect their true values\footnote{The ``true'' values of $\mathcal{H_O}$ etc. refers to their actual values when calculated directly through their definitions. These values represent the actual/true values of the parameters along the individual light rays {\em in the given model including grid and interpolation choice}.} for the given gridded spacetime. This is also illustrated in table \ref{table:fit} which includes a comparison of the true values of $\mathcal{H_O}$ and $\mathcal{Q_O}$ with the values obtained from fitting a third order polynomial to the corresponding redshift-distance data. From this fit, $\mathcal{H_O}$ is obtained as $c_1/c$, and $\mathcal{Q_O}$ by $1 - 2\mathcal{H_O}c_2/c$. As the table shows, the fitted values deviate significantly from the true ones, especially $\mathcal{Q_O}$ which depends on higher order derivatives. $\mathcal{H_O}$ is stable at percent order when going between the cubic and Steffen interpolation schemes with N = 8 and in these cases the fitted value of $\mathcal{H_O}$ is also very close to the true values. For the N = 4 grid, both $\mathcal{H_O}$ and $\mathcal{Q_O}$ are similar when comparing the true and fitted values along a given line of sight and ``merely'' deviate from each other at percent level. Most importantly, however, the table shows that the fitted coefficients change as the grid is smoothed. As noted above, the fitted coefficients correspond to the results one would obtain from fitting real-world observations to the cosmographic expansion. In \cite{gevolutiondipole}, such fits were made to synthetic data from a simulation and it was suggested that the fitted coefficients could be interpreted as those corresponding to a smoothed spacetime. The results presented here emphasize that it is not clear that one can fit a third order polynomial to data of arbitrary detail and consider the fitted coefficients as representing the cosmographic coefficients of a corresponding smoothed spacetime. Such an interpretation becomes even more problematic when remembering that real data does not typically yield the redshift-distance relation along individual light rays but rather represents discrete points, each corresponding to a different light ray. This is discussed further in section \ref{sec:summary}.

\begin{figure}
    \centering
    \includegraphics[width=\columnwidth]{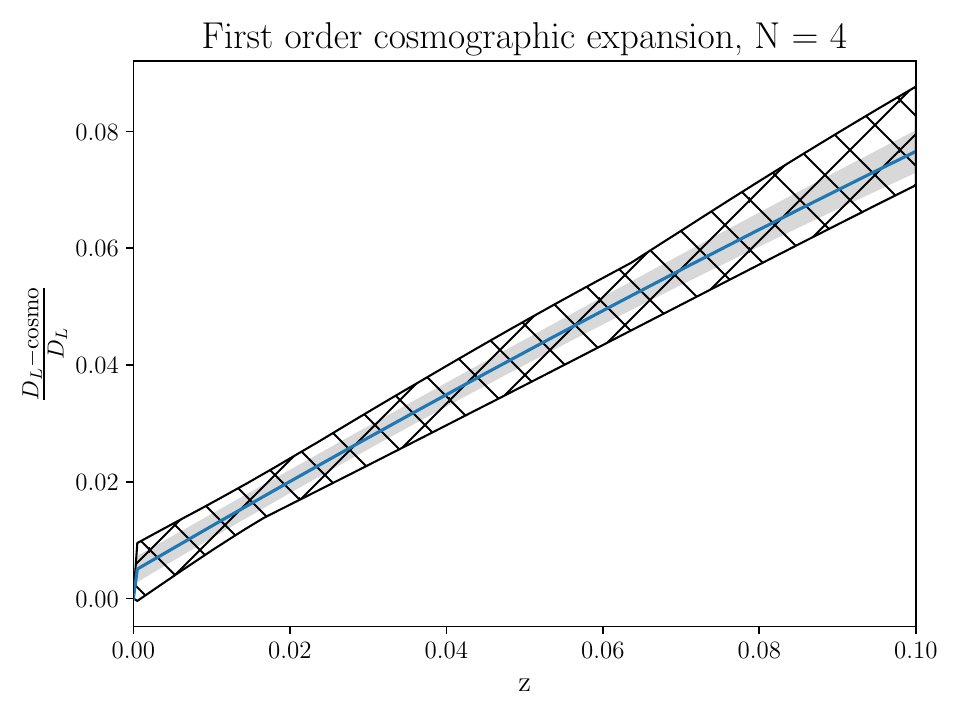} \\
    \includegraphics[width=\columnwidth]{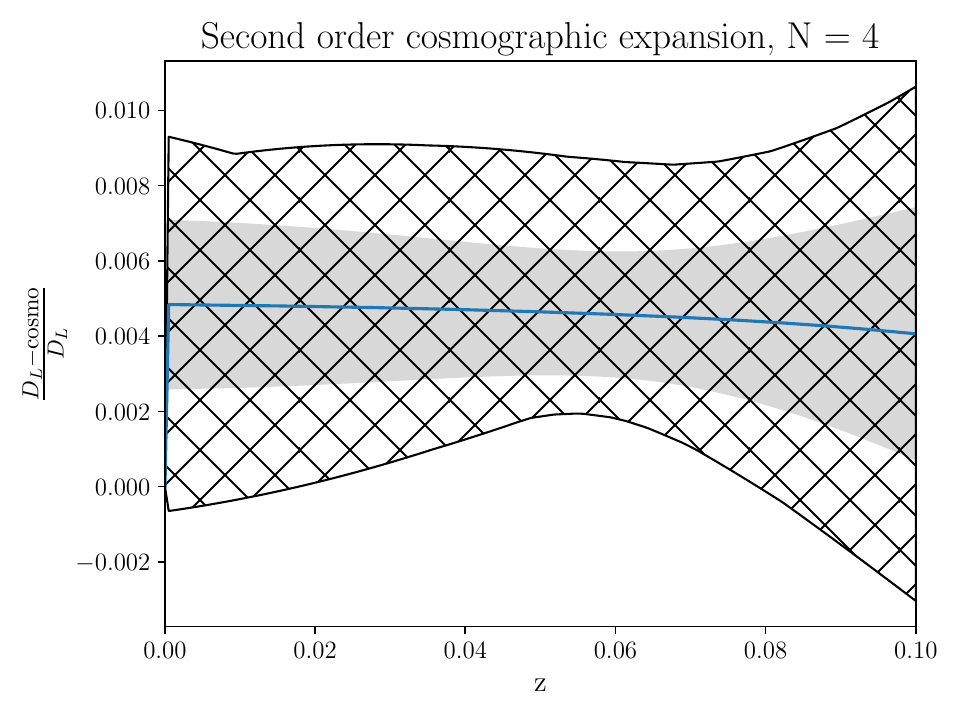}\\
       \includegraphics[width=\columnwidth]{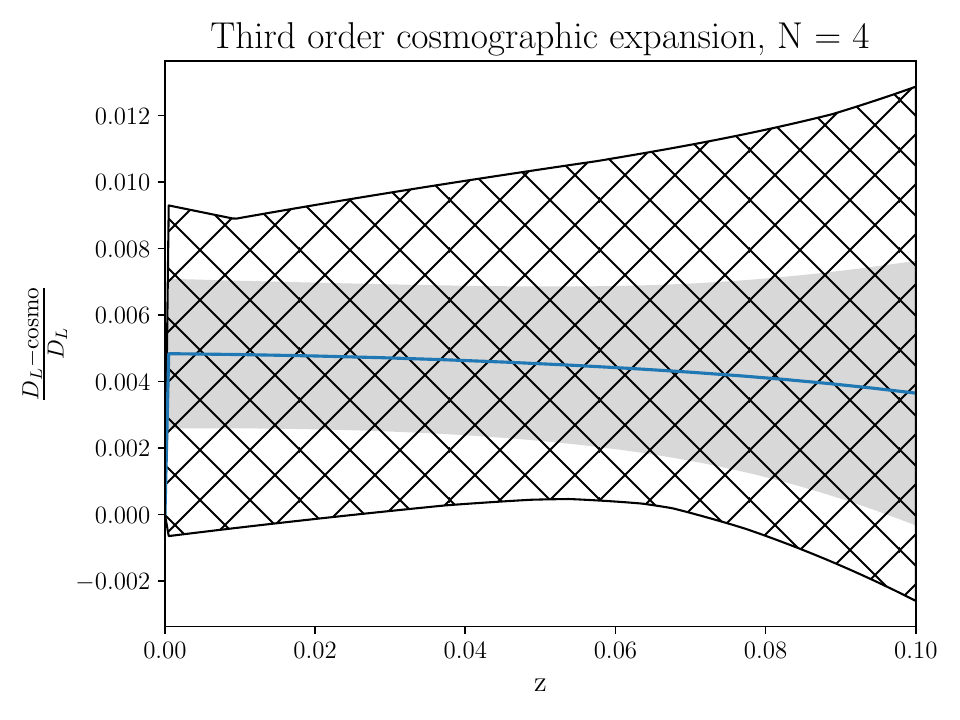}
    \caption{Relative deviation between the exact luminosity distance ($D_L$) and the cosmographic expansion (``cosmo'') at first, second and third order using cubic interpolation on the N = 4 grid. Hatched areas show the fluctuation along all light rays while the shaded area shows a standard deviation around the mean which is also plotted.
    }
    \label{fig:dcososmo_z_grid4}
\end{figure}

\begin{figure}
    \centering
    \includegraphics[width=\columnwidth]{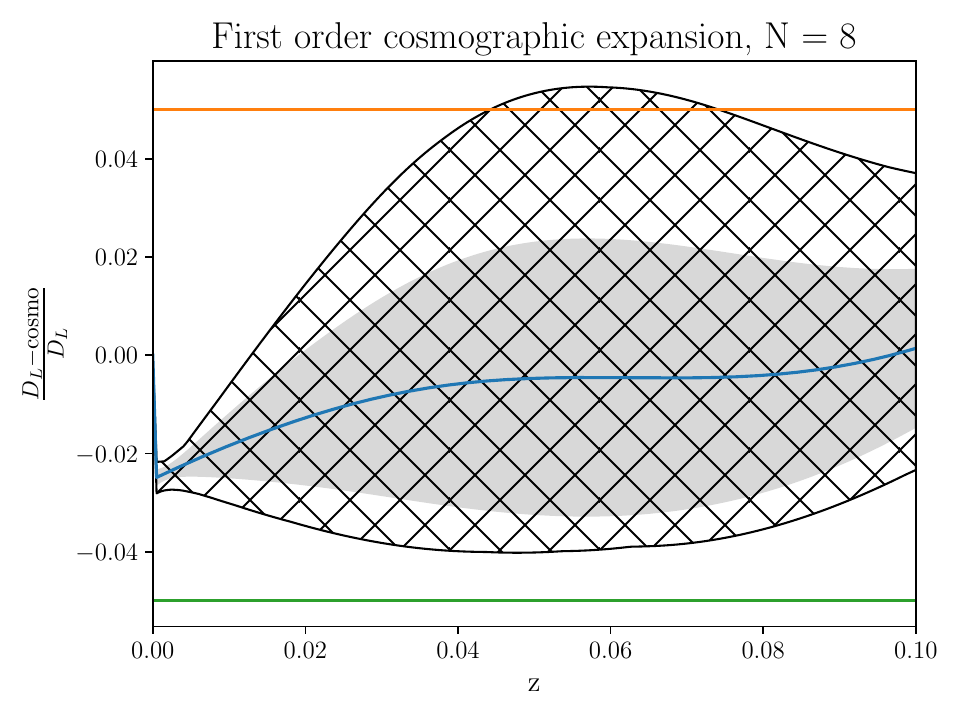} \\
    \includegraphics[width=\columnwidth]{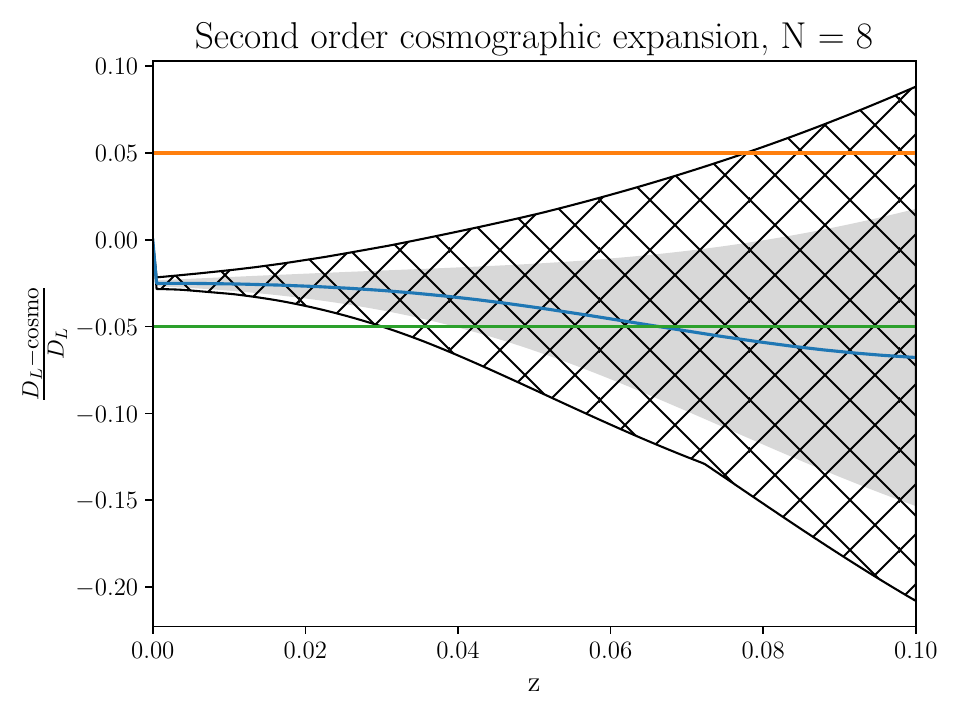}\\
       \includegraphics[width=\columnwidth]{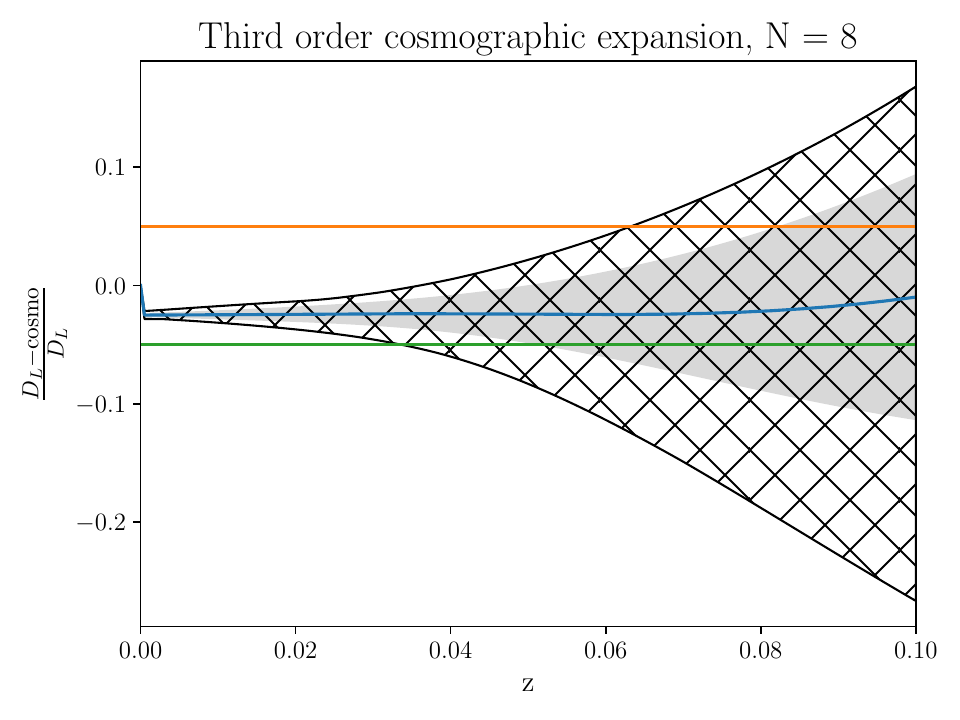}
    \caption{Relative deviation between the exact luminosity distance ($D_L$) and the cosmographic expansion (``cosmo'') at first, second and third order using cubic interpolation on the N = 8 grid. Hatched areas show the fluctuation along all light rays while the shaded area shows a standard deviation around the mean which is also plotted. To ease reading of the plots, lines are shown at $\pm 0.05$.
    }
    \label{fig:dcososmo_z_grid8}
\end{figure}

\subsection{Multiple light rays}
The previous results overall show that the cosmographic expansions are very ``unstable'': The coefficients depend strongly on the interpolation scheme, grid resolution and grouping of data, and are not generally well approximated by fitted coefficients. However, for the lowest grid resolution considered (N = 4) the ``true'' values of $\mathcal{H_O}$ and $\mathcal{Q_O}$ are fairly stable against changes in the interpolation scheme and for N = 8 this is also true, although only for $\mathcal{H_O}$. In this section we will therefore consider how these quantities vary across the sky. For this, 768 light rays are in each of these models distributed across the observed sky using Healpix\footnote{https://healpix.sourceforge.io/}. We will consider only the ungrouped data here as the initial study based on ray 1 and ray 2 indicates that the results are very similar in the grouped versus ungrouped models. This is also the conclusion when considering 768 lines of sight for the simplest model with linear interpolation and N = 64 for which results are shown later in this section.
\newline\newline
Figure \ref{fig:dcososmo_z_grid4} shows the relative deviation between the exact redshift-distance relation and the cosmographic expansion at first, second and third order for the model based on cubic interpolation on the N = 4 grid. The second order cosmographic expansion makes a visibly better reproduction of the exact redshift-distance relation while little is changed when going to third order. This is presumably because the grid is smoothed to 250Mpc/h and thus higher order derivatives entering into the third order coefficient are small. If considering only one standard deviation, the second and third order cosmographic expansions reproduce the exact redshift-distance relation to within one percent.
\newline\indent
\begin{figure*}
    \centering
    \includegraphics[width=\columnwidth]{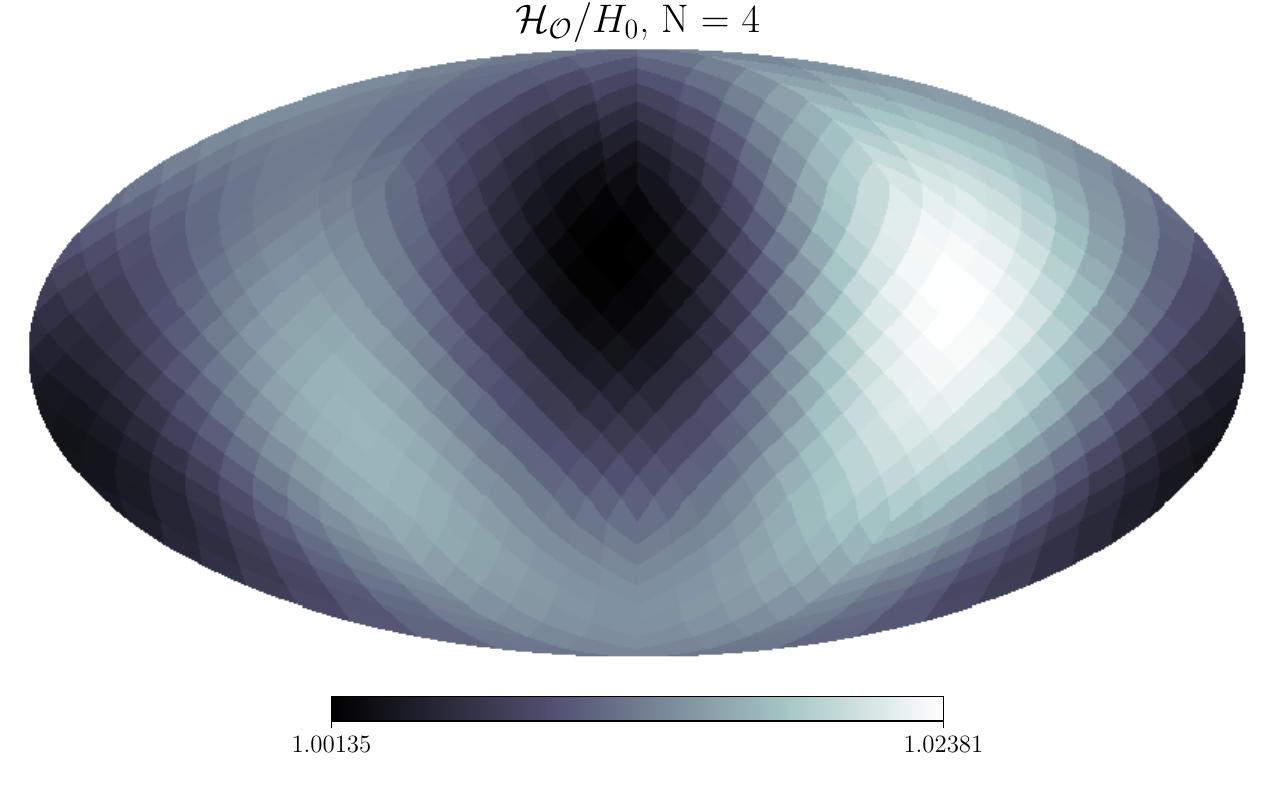}
    \includegraphics[width=\columnwidth]{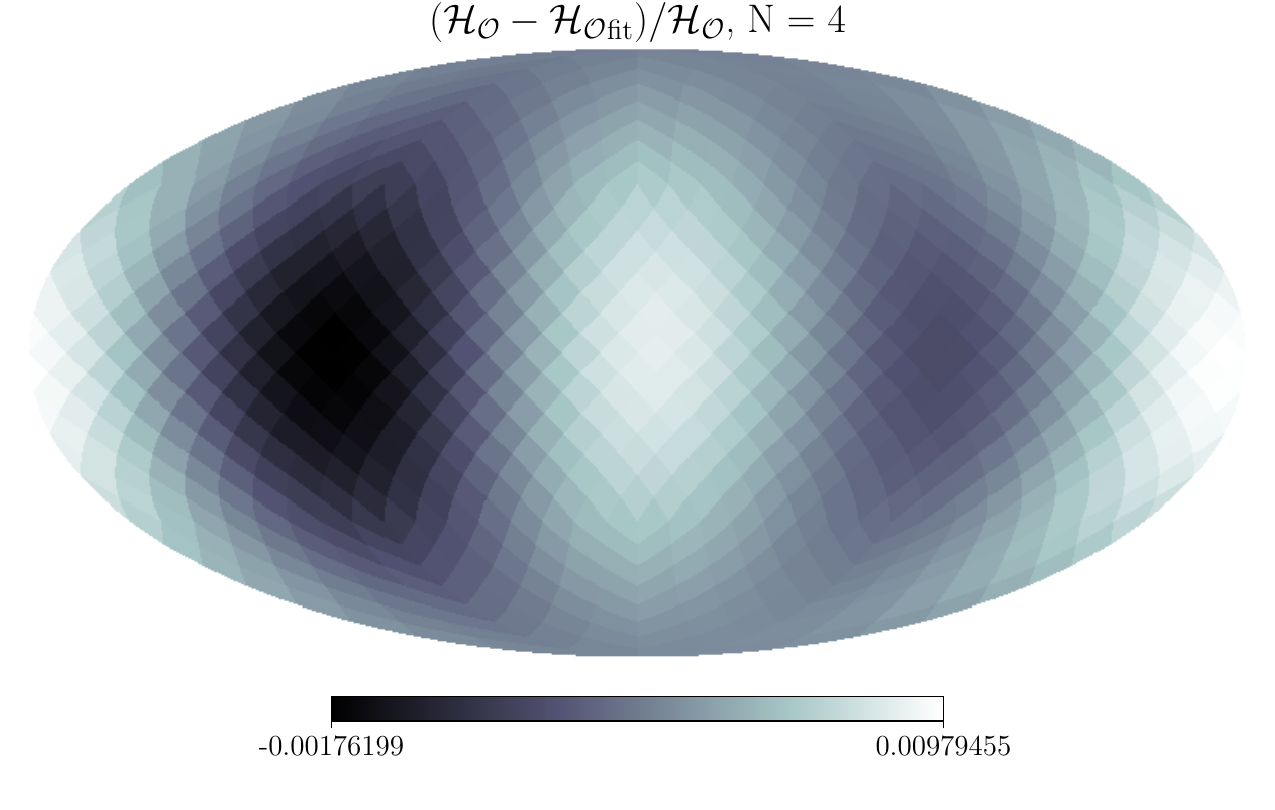}\\
        \includegraphics[width=\columnwidth]{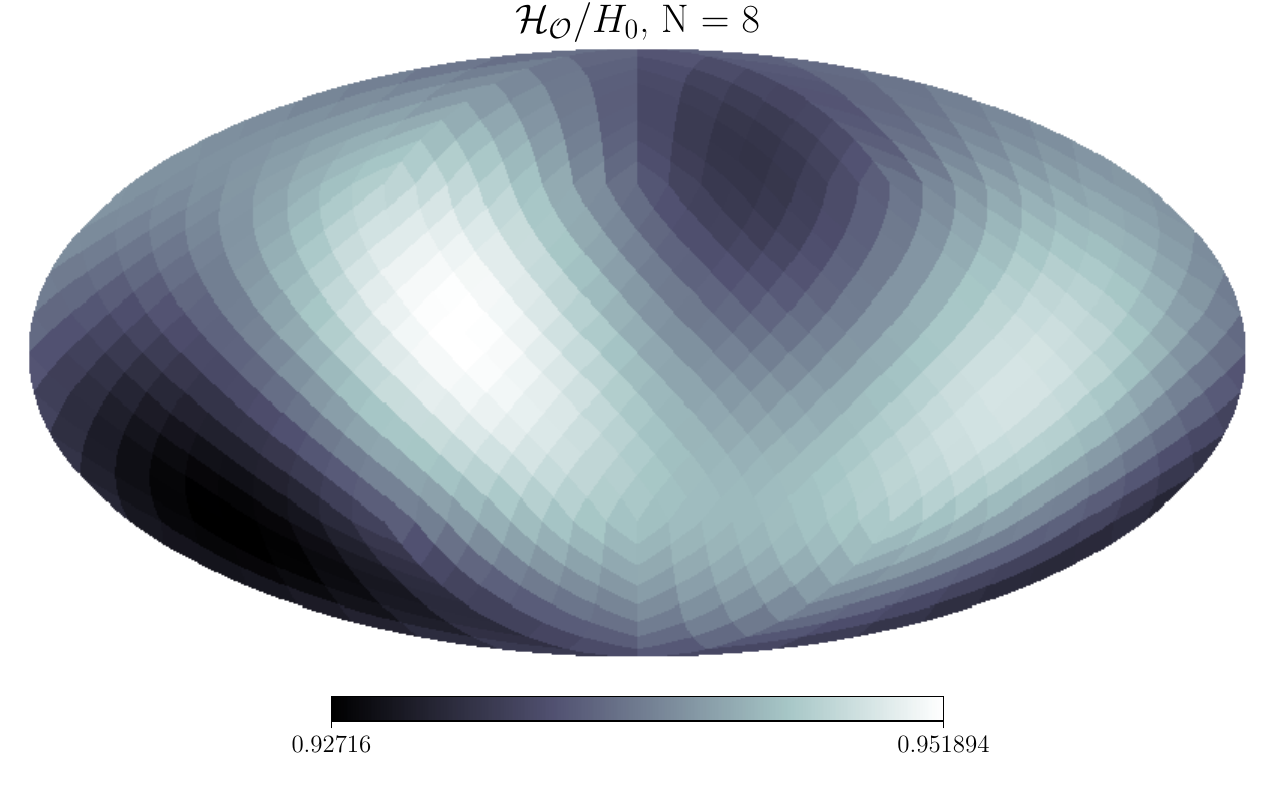}
    \includegraphics[width=\columnwidth]{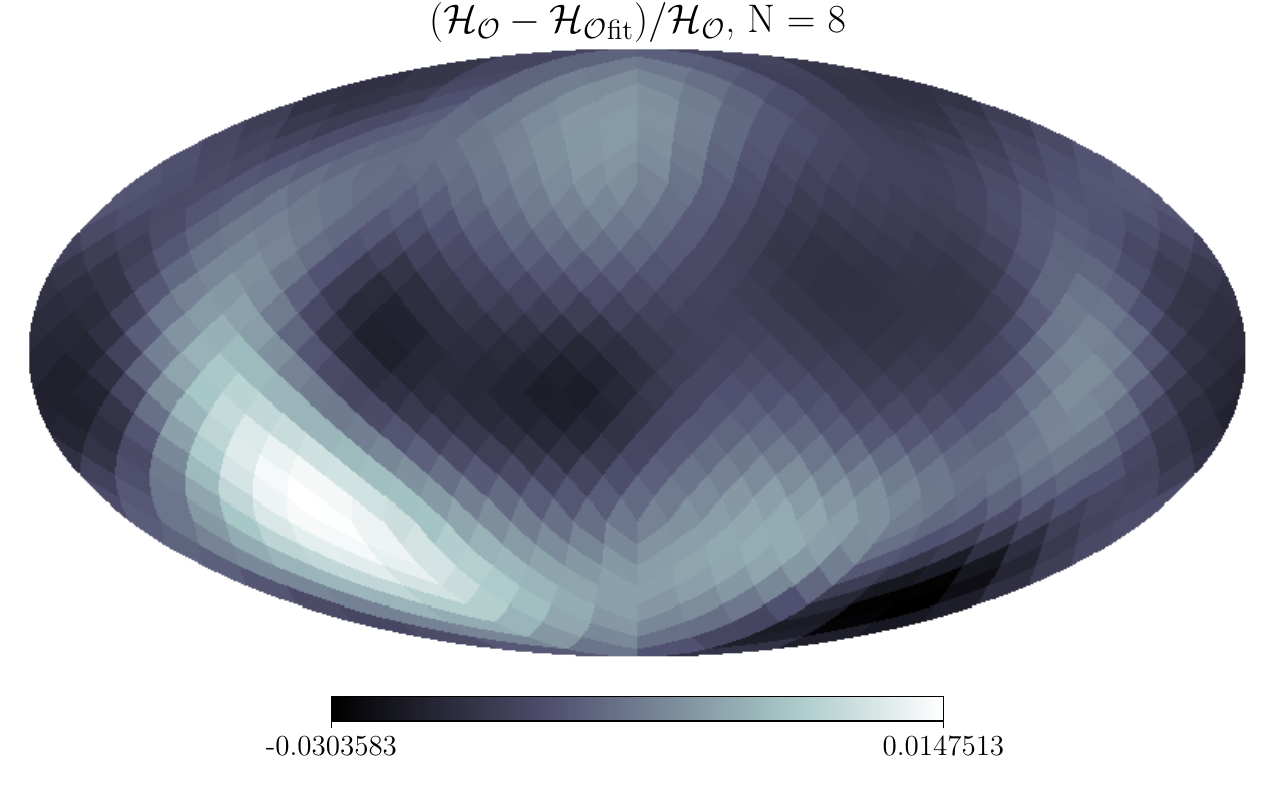}
    \caption{All-sky maps of $\mathcal{H_O}$ for the considered observer, depicted relative to the background value of $H_0$. The sky maps are shown for the models based on cubic interpolation and the grids N = 4 and N = 8. The plots show both the exact/true $\mathcal{H_O}$ and the deviation between its true and fitted values.}
    \label{fig:curlH_grid4and8}
\end{figure*}

\begin{figure*}
    \centering
    \includegraphics[width=\columnwidth]{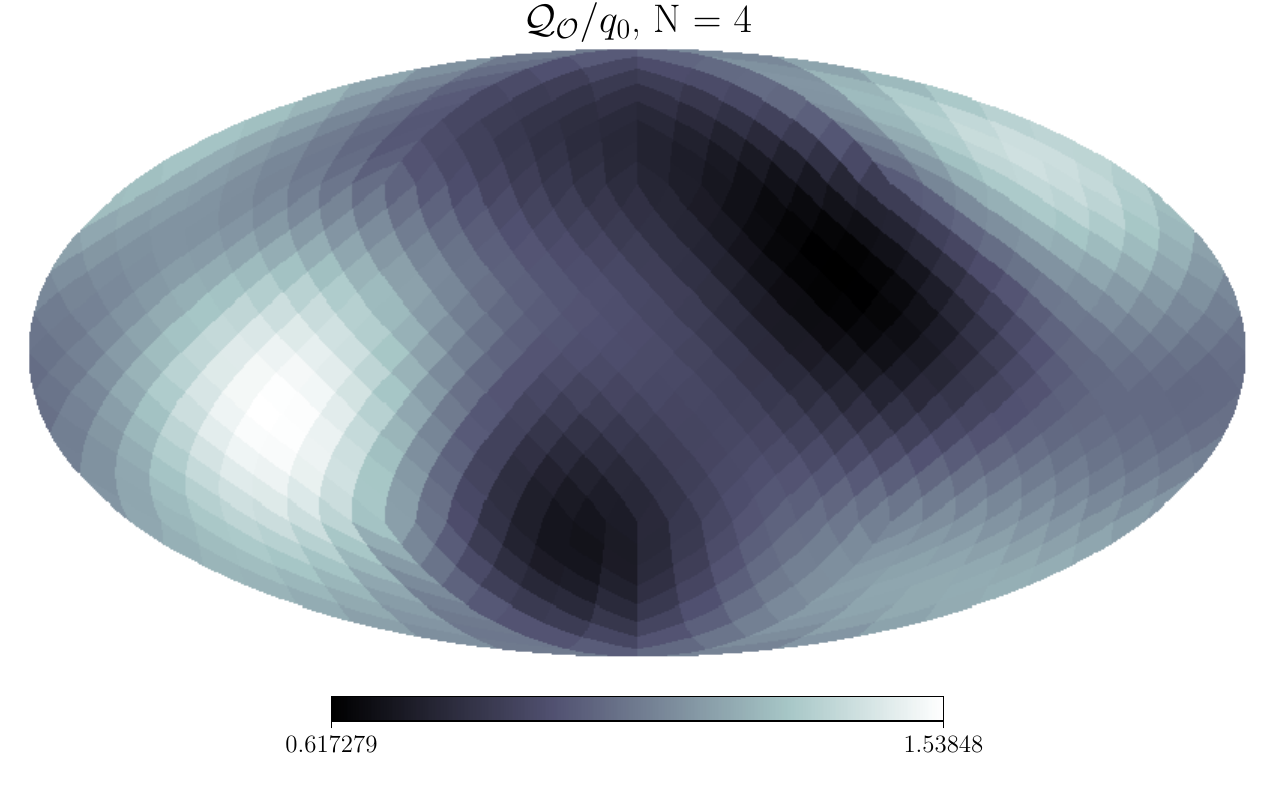}
    \includegraphics[width=\columnwidth]{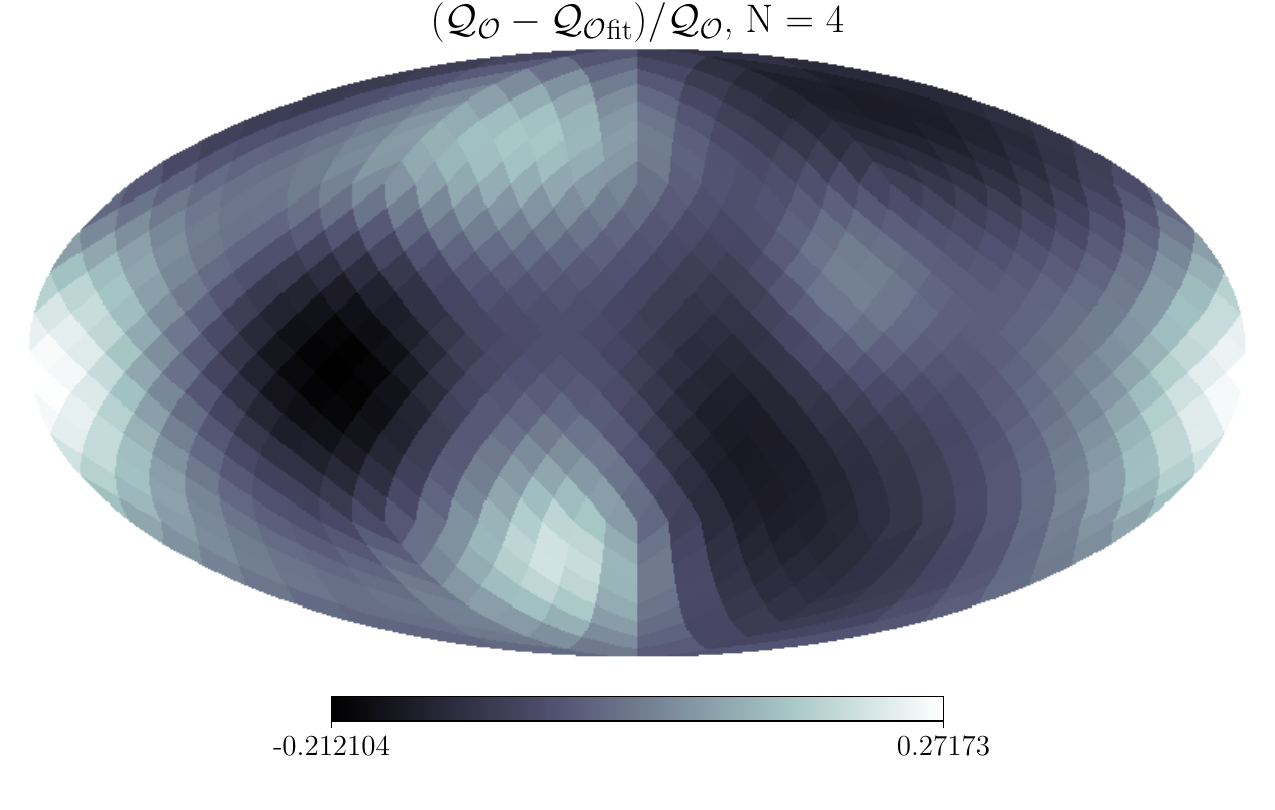}\\
    \includegraphics[width=\columnwidth]{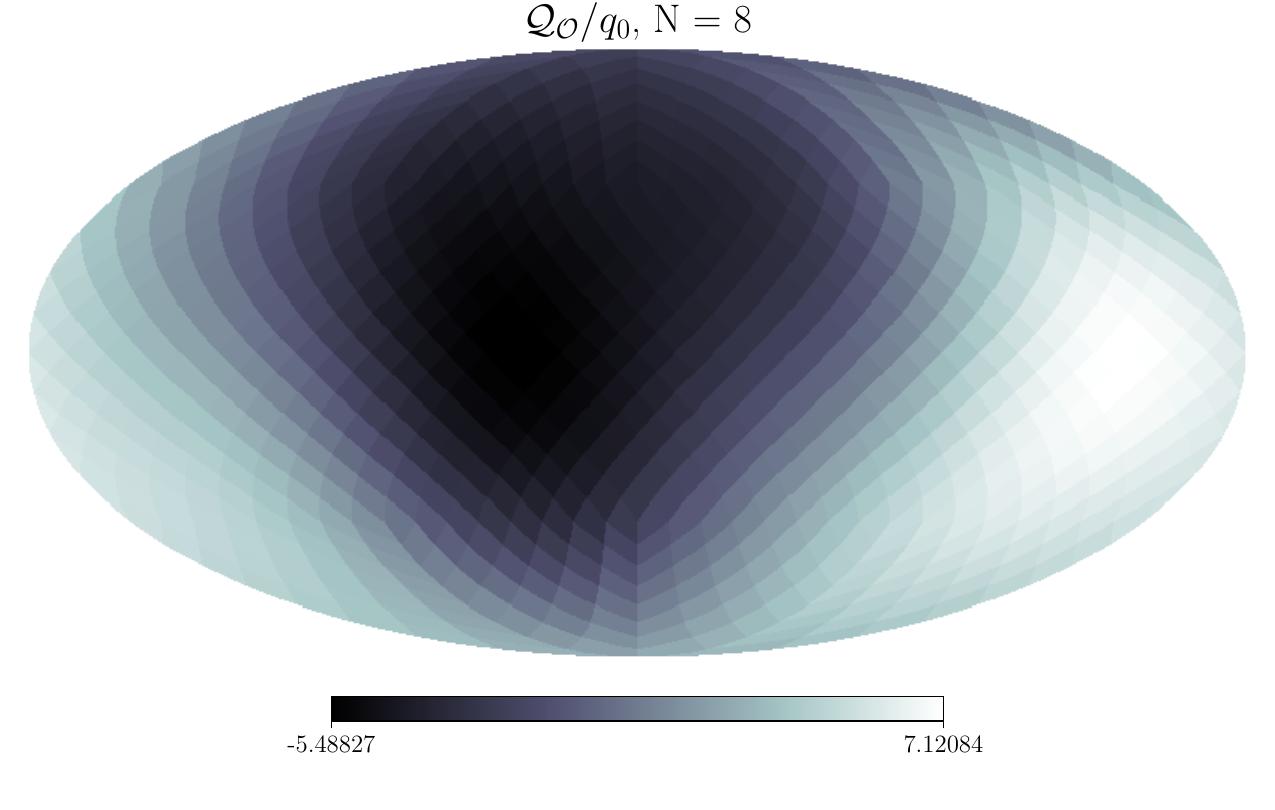}
    \includegraphics[width=\columnwidth]{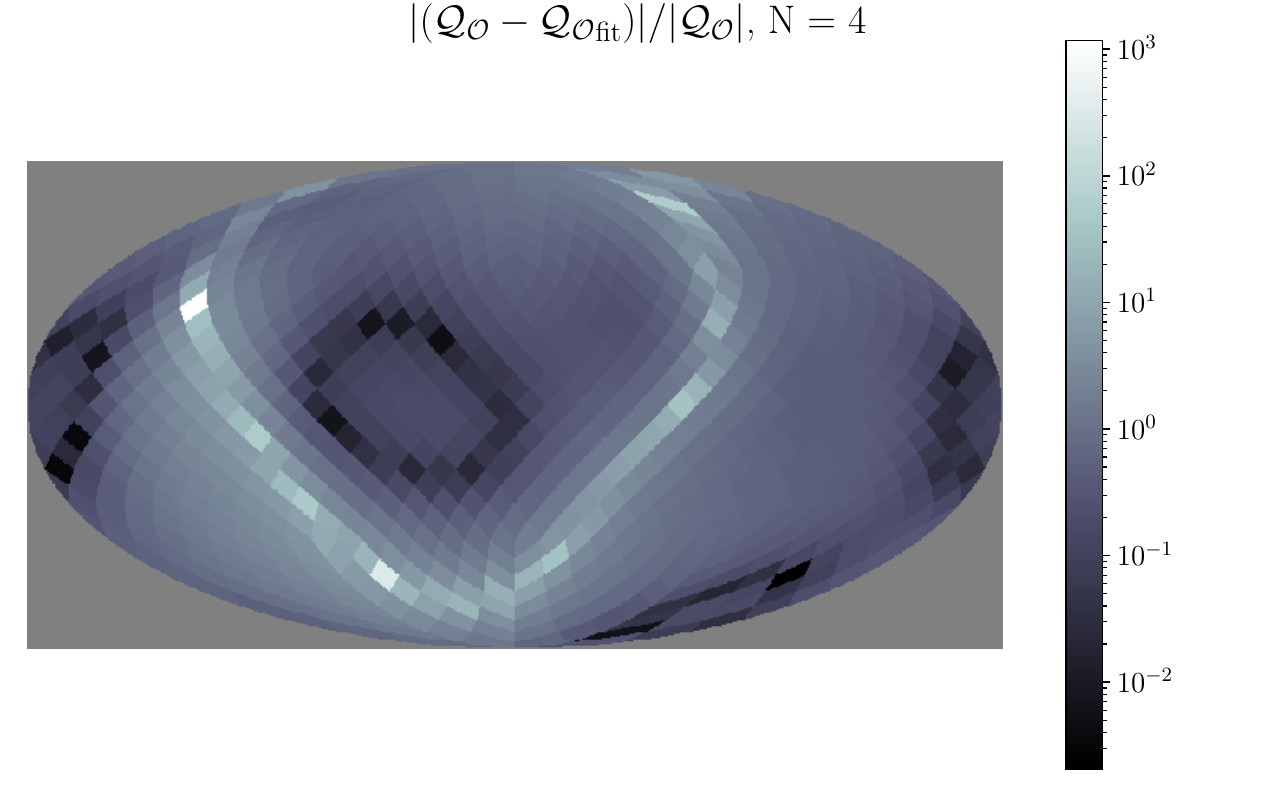}
    \caption{All-sky maps of $\mathcal{Q_O}$ for the considered observer, depicted relative to the background value, $q_0$. The sky maps are shown for the models based on cubic interpolation and the grids N = 4 and N = 8. The plots show both the true $\mathcal{Q_O}$ and the deviation between its true and fitted values. Note that for N = 8, the plot showing the deviation between true and fitted values is shown on a log-scale.}
    \label{fig:curlQ_grid4and8}
\end{figure*}
\begin{figure}
    \centering
    \includegraphics[width=\columnwidth]{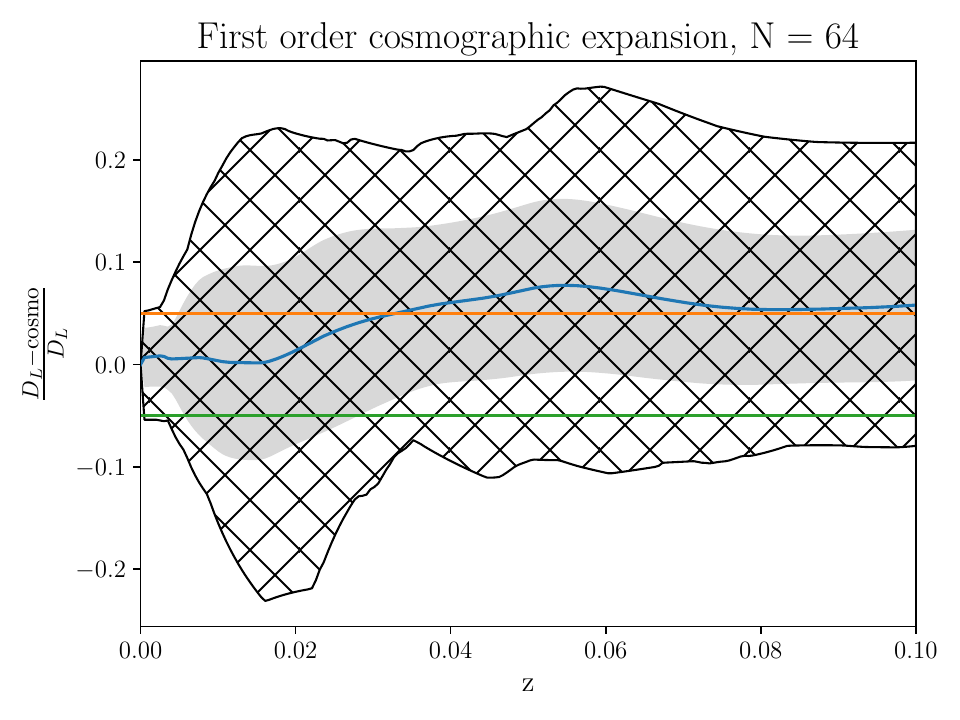} 
    \includegraphics[width=\columnwidth]{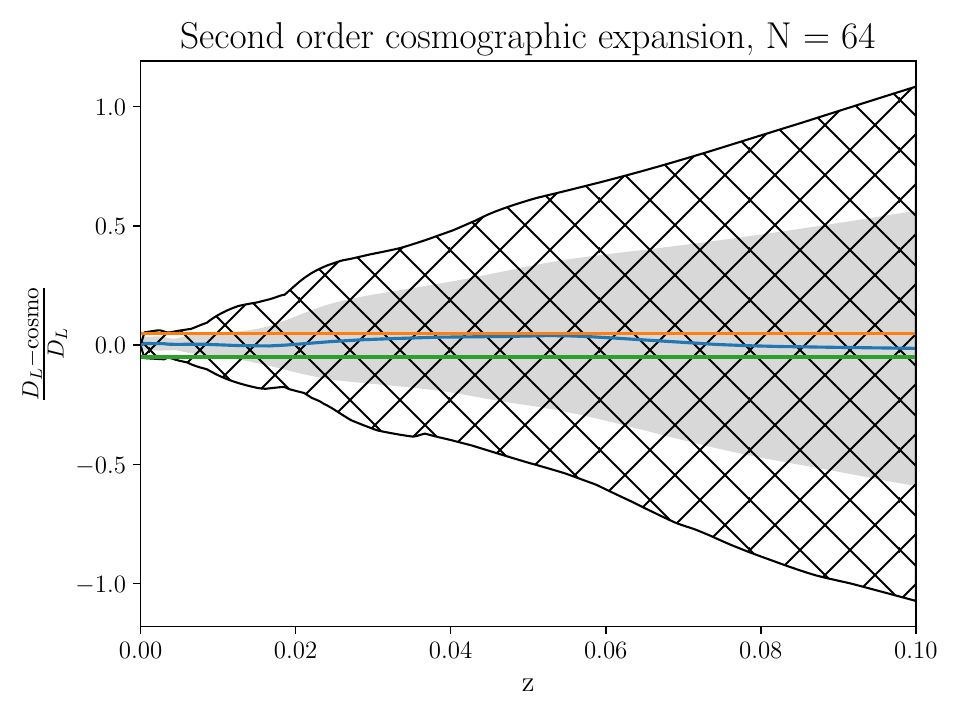}
       \includegraphics[width=\columnwidth]{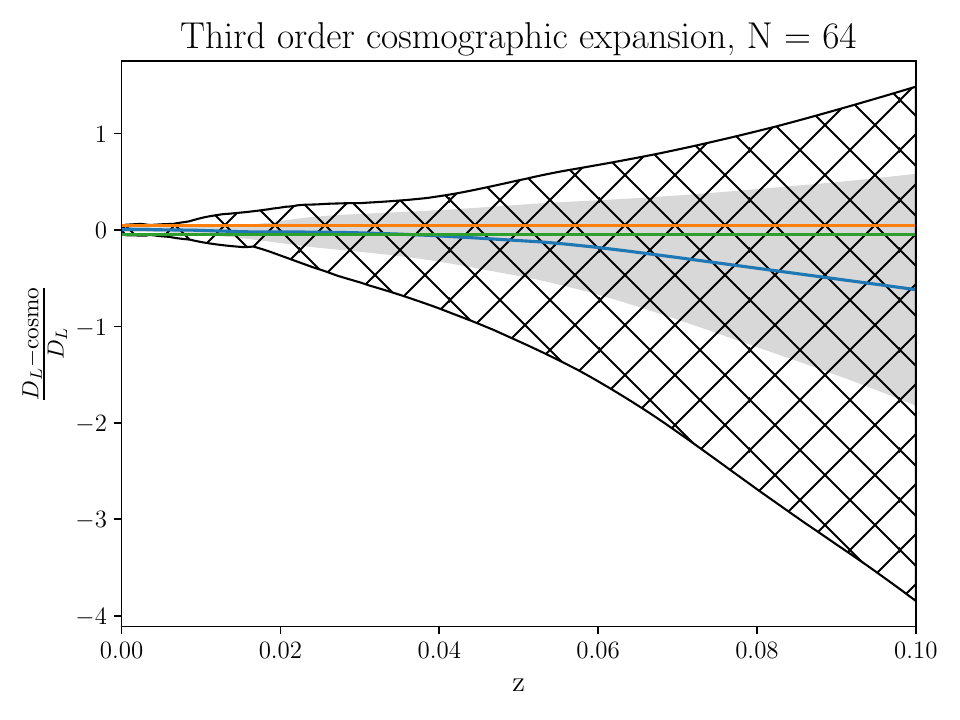}
    \caption{Relative deviation between the exact luminosity distance ($D_L$) and the cosmographic expansion (``cosmo'') at first, second and third order for the model based on linear interpolation and N = 64. Hatched areas show the fluctuation along all light rays while the shaded area shows a standard deviation around the mean which is also plotted and appears as the slightly curving line in the plots. To ease interpreting the plots, the figures include lines at $\pm 5\%$.}
    \label{fig:dcosmo_z}
\end{figure}

The relative deviation between the exact redshift-distance relation and cosmographic expansions for the model based on cubic interpolation and N = 8 are shown in figure \ref{fig:dcososmo_z_grid8}. At second and third order, the cosmographic expansion is clearly poorer for the N = 8 grid than for the n = 4 grid, and divergence of the expansion from the exact redshift-distance relation begins much before $z = 0.1$ is reached. Already at $z \approx 0.03$, the second and third order expansions becomes incorrect above 5\%. The third order expansion is also clearly seen to be {\em worse} than the second order expansion (except possibly at very low redshifts), again indicating that the Taylor expansion is not converging.
\newline\newline
Figure \ref{fig:curlH_grid4and8} shows the generalized Hubble parameter, $\mathcal{H_O}$, for the model based on cubic interpolation and the grids N = 4, 8. The fluctuations are shown relative to the background Hubble constant, $H_0 = 74.6$km/s/Mpc. For the grid = 4 model, the grid has been smoothed to such an extent that the generalized expansion rate fluctuates by only a few percent across the entire sky. In this case, the values of $\mathcal{H_O}$ obtained from fitting to a polynomial (with the form of $p(z)$) corresponds very well with the true value. For the N = 8 grid, $\mathcal{H_O}$ fluctuates significantly more across the sky, although still below 10\%. In this case, the deviation between the true and fitted values of $\mathcal{H_O}$ reaches about a factor of 30 higher than in the N = 4 case, and the deviations now reach a few percent at maximum. As a side remark, one may note that the values of $\mathcal{H_O}$ are consistently above the background value for the N = 4 grid while it is consistently below the background value for N = 8. This is an important reminder that smoothing has a significant impact on the results, especially when gradients are being calculated.
\newline\indent
Figure \ref{fig:curlQ_grid4and8} shows $\mathcal{Q_O}$ across the sky, depicted relative to the background deceleration parameter $-0.55$. The fluctuations in $\mathcal{Q_O}$ are much stronger than in $\mathcal{H_O}$ and reach over 50\% in the most extreme patches of the sky even for the N = 4 grid. For this parameter, even for the N = 4 grid, the fitted values deviate by up to around 30\% compared to the true value. For the N = 8 grid, $\mathcal{Q_O}$ fluctuates significantly more across the sky, reaching several hundred percent. In this case, the true and fitted values of $\mathcal{Q_O}$ deviate from each other more than a factor of 10 above the deviation found in the N = 4 case. When assessing the importance of this result, it should be remembered that the N = 8 grid, the value of $\mathcal{Q_O}$ depends significantly on the chosen interpolation scheme.
\newline\newline
Lastly, we will consider the results obtained when using the naive model based on using linear interpolation together with the N = 64 grid for comparison with the results obtained using N = 4 and N = 8.
\newline\indent
Figure \ref{fig:dcosmo_z} shows the deviations between the exact luminosity distance versus the first, second and third order cosmographic expansions along all lines of sight for the ungrouped data. The results have been reproduced using grouped data. Since the results obtained using grouped and ungrouped data are very similar, the results obtained with the grouped data are not shown. The maximum fluctuations are quite large (above $5-10\%$) already at very low redshifts $z\sim 0.005-0.01$. Only considering one standard deviation around the mean (very) roughly halves the amplitude of the scatter. In addition, as already discussed when considering individual light rays above, the precision of the numerical investigation is quite low and a 5\% error should not be considered particularly problematic as it can simply be due to using the finite difference on a very crude grid combined with linear interpolation. The important results appear when considering the second and third order expansions where the deviations from the exact redshift-distance relation increase very rapidly and become of order $50 - 100\%$ before $z \approx 0.05$. These values are approximately halved when only considering one standard deviation around the mean. For the third order cosmographic expansion, the mean even seems to diverge, indicating that the lines of sight with divergent cosmographic expansions dominate the statistical results. This is in line with the skewness seen in the figure which shows that the divergence typically is such that the cosmographic expansion is larger than the exact luminosity distance.
\newline\newline

\begin{figure}
    \centering
    \includegraphics[width=\columnwidth]{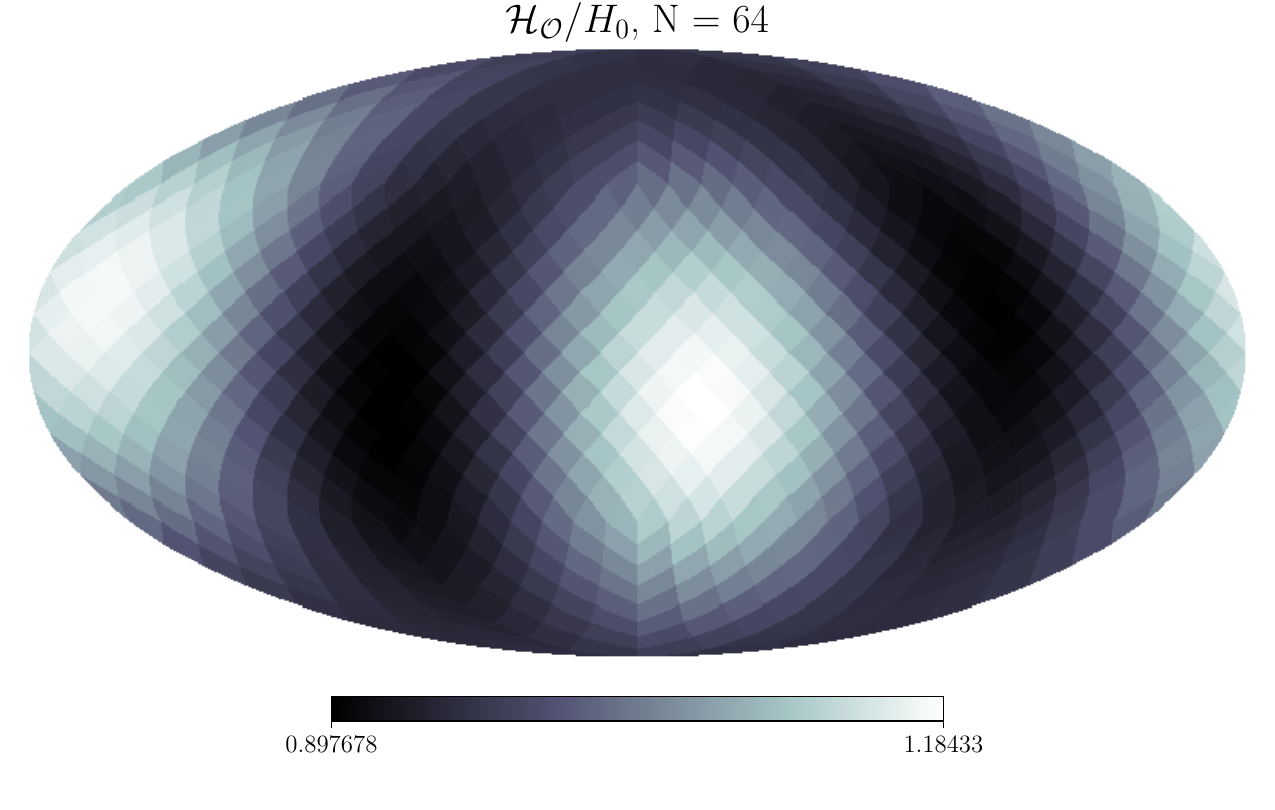}
    \caption{All-sky map of $\mathcal{H_O}$ for the considered observer, depicted relative to the background value of $H_0$.}
    \label{fig:curlH}
\end{figure}
\begin{figure}
    \centering
    \includegraphics[width=\columnwidth]{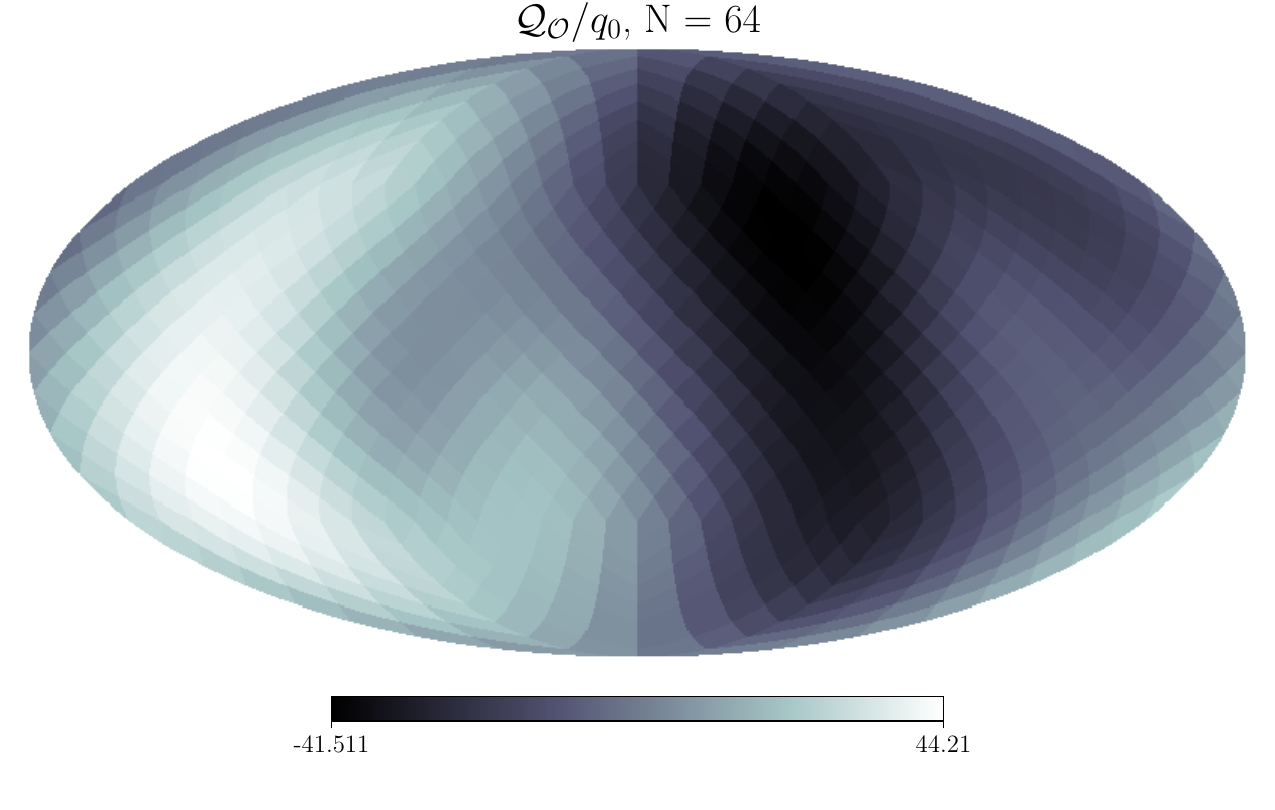}
    \caption{All-sky map of $\mathcal{Q_O}$ for the considered observer, depicted relative to the background value, $q_0$.}
    \label{fig:curlQ}
\end{figure}

The main point of figure \ref{fig:dcosmo_z} is to emphasize that the cosmographic expansion does not in general converge unless the inhomogeneities are smoothed very strongly as in the N = 8 and N = 4 cases. Even so, it is interesting to see how significant the fluctuations are in the exact $\mathcal{H_O}$ and $\mathcal{Q_O}$ when using the naive model based on linear interpolation and N = 64. Figure \ref{fig:curlH} therefore shows the all-sky distributions of $\mathcal{H_O}$ for the model based on linear interpolation and N = 64. We see that $\mathcal{H_O}$ reaches up to approximately $18\%$ larger than the background value and $11\%$ smaller along the most extreme lines of sight (for the model based on the grouped data, these fluctuations reduce to roughly $12\%$ and $10\%$, respectively). Remember here, that when using N = 64, the results depend strongly on the chosen interpolation scheme. The main point therefore simply is that, qualitatively speaking, the fluctuations in $\mathcal{H_O}$ are significantly larger here than for the N = 4 and N = 8 grids considered earlier.
\newline\indent
Figure \ref{fig:curlQ} shows the all-sky distributions of $\mathcal{Q_O}$ for the model. As seen, the values along the 768 lines of sight fluctuate around the background value by approximately a factor of $\pm40$. For the model based on the grouped data, the fluctuations are only slightly smaller (not shown).


\subsection{Comparison to earlier work}\label{subsec:compare}
\begin{figure}
    \centering
    \includegraphics[width=\columnwidth]{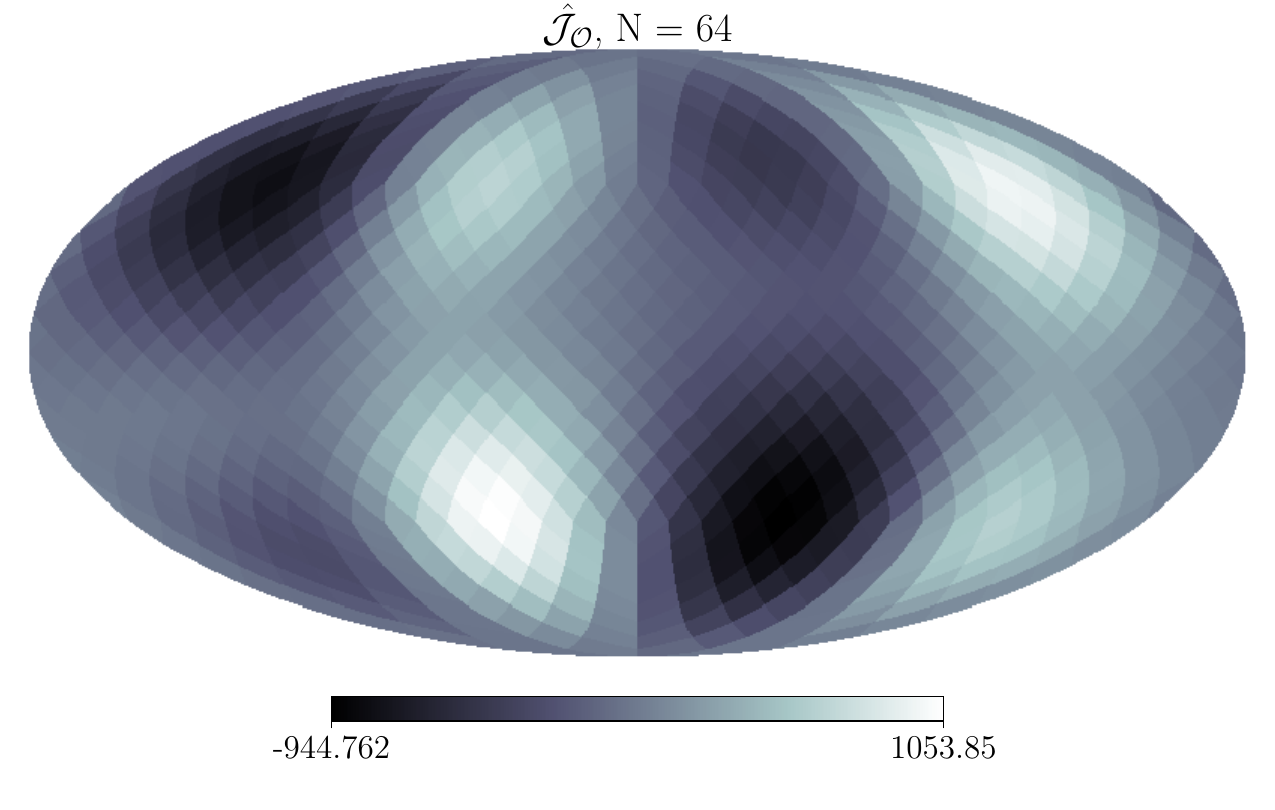}\\
        \includegraphics[width=\columnwidth]{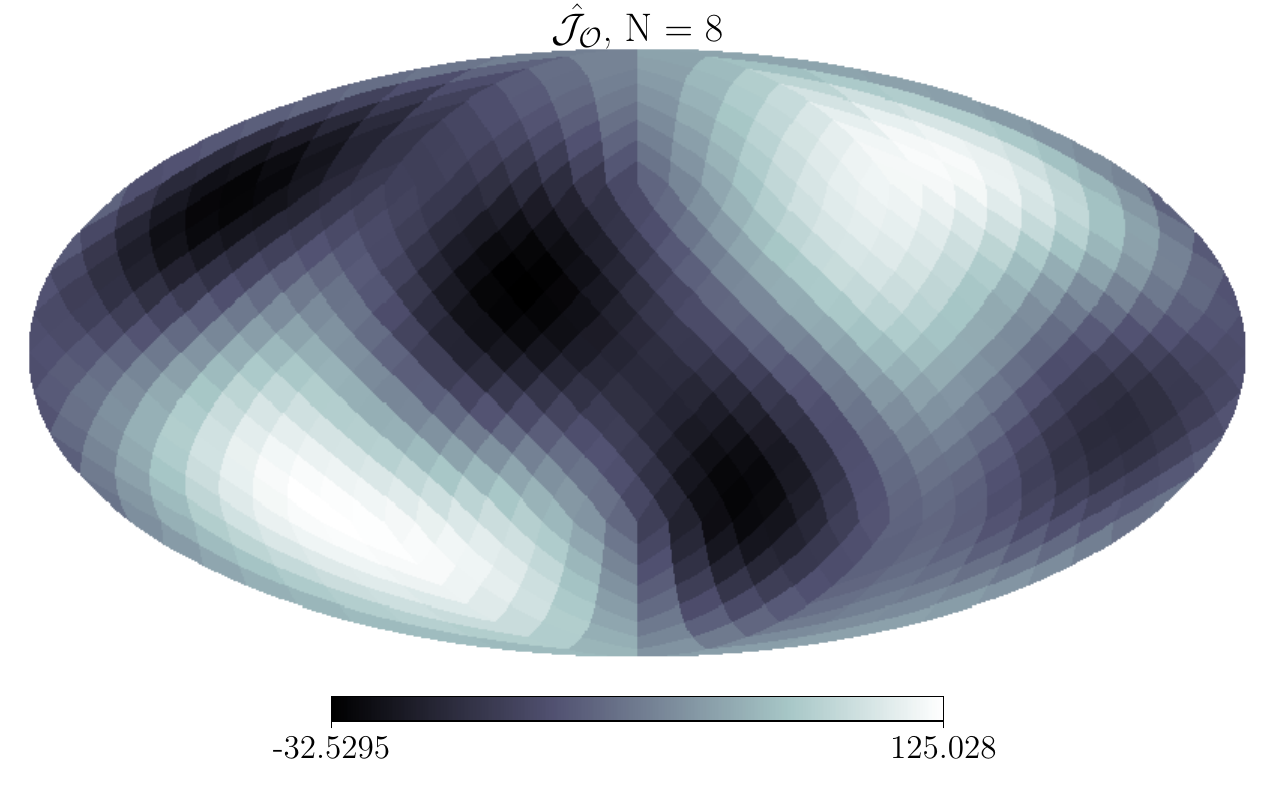}\\
            \includegraphics[width=\columnwidth]{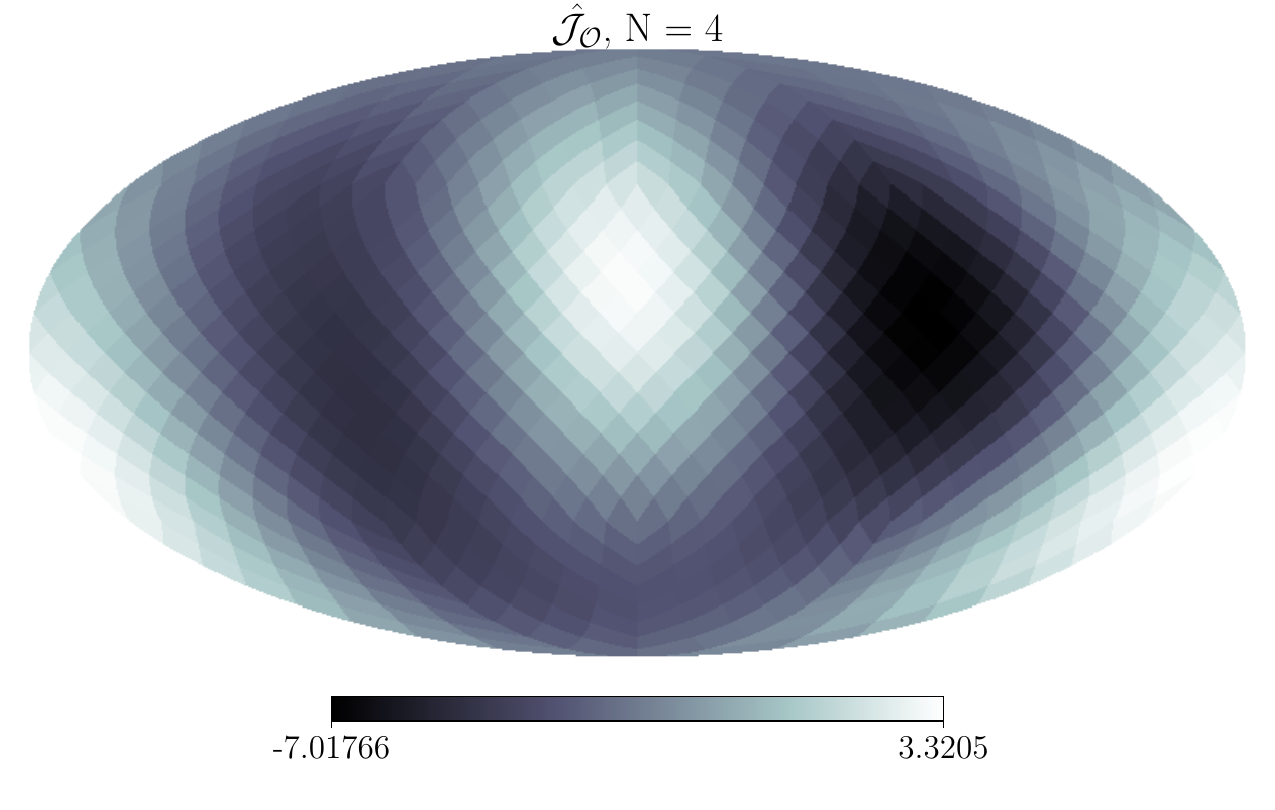}
    \caption{All-sky maps of $\hat{\mathcal{J_O}}$ for the considered observer. The sky maps are shown for the models based on linear interpolation with N = 64, and cubic interpolation with N = 4 and 8.}
    \label{fig:curlJhat}
\end{figure}

\begin{figure}
    \centering
\includegraphics[width=\columnwidth]{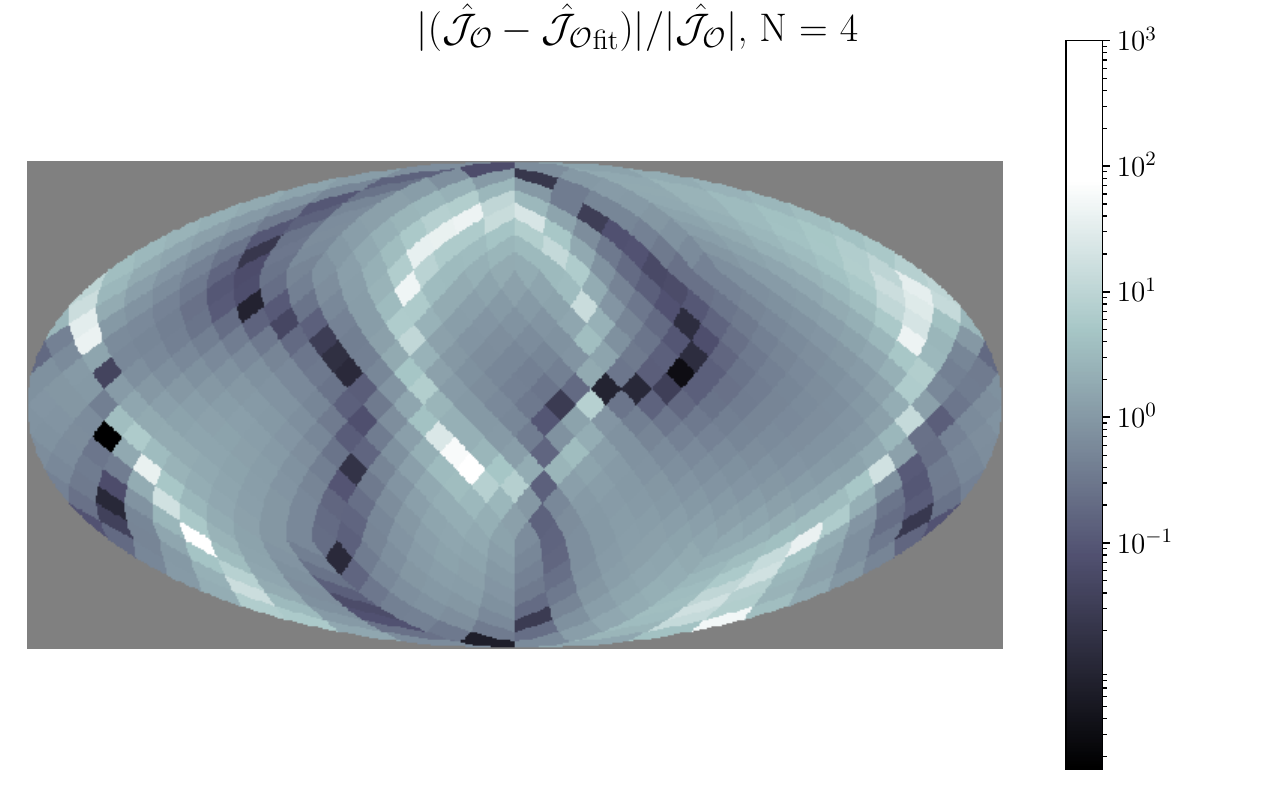}
    \caption{All-sky map of deviation between true and fitted values of $\hat{\mathcal{J_O}}$ for the model based on cubic interpolation with N = 4. The color bar is logarithmic.}
    \label{fig:curlJhat_fit}
\end{figure}

In \cite{ETdipole1}, the general cosmographic expansion was used in low-resolution simulations with a density contrast amplitude of $\sim 0.05$ made with the Einstein Toolkit. The authors found that even in a simulation with such small density contrast, random observers see sky-variance of $\mathcal{H_O}$ of typically $2\%$ and of $\mathcal{Q_O}$ of roughly 120\%. The fluctuations found here for $\mathcal{H_O}$ are about a factor of 5-10 higher when using the N = 64 grid. For $\mathcal{Q_O}$, the all-sky fluctuations in $\mathcal{Q_O}$ have a maximum amplitude about of the order of 4000 percent around the background value of $-0.55$. This is significantly more than the fluctuations found in \cite{ETdipole1}, but considering that higher density contrasts are used here, this is not particularly troublesome. When smoothing the resolution and using N = 4, 8 we find fluctuations in $\mathcal{H_O}$ similar to those found in \cite{ETdipole1}.
\newline\newline
In \cite{gevolutiondipole}, the general cosmographic expansion was recast using the Padé approximant \cite{pade_summary} and used in combination with simulations made with gevolution and a randomly placed observer. The main simulation studied had a resolution of 1.75Mpc/h \cite{resolution}, but the study included a comparison with a ``smoothed'' version of the simulation. From figures 6-8 in \cite{gevolutiondipole} it is seen that the components of the expansions can overall only be constrained accurately/correctly when considering very low redshifts and/or using the smoothed version of the simulation. The only somewhat consistently good results are obtained for the smoothed simulation using data only at $z<0.025$. This is in line with the results of \cite{ETdipole2}, where it was found that even in a smoothed simulation made with the Einstein Toolkit (equivalent to the smoothed simulation used in \cite{gevolutiondipole}), the cosmographic expansion (to third order) was inaccurate above $\sim 10\%$ for $z>0.04$. The results of \cite{ETdipole2} also demonstrate that the accuracy of the cosmographic expansion decreases significantly when the simulations include successively more structure. These results are overall in agreement with the results found here, where the cosmographic expansion is found to break down along a significant portion of the lines of sight long before $z = 0.1$ is reached when considering the model with N = 64 and down to N = 8. Only the models based on N = 4 have redshift-distance relations reproduced accurately by the cosmographic expansions up to $z\approx 0.1$.
\newline\indent
It is worth noting that the results obtained here, in \cite{gevolutiondipole} and in \cite{ETdipole1, ETdipole2} are based on different approximations. The approximations made in the current work include the weak field assumption, setting $v_{,t} = 0$, and neglecting perturbations to the light paths. These approximations were are also included in \cite{gevolutiondipole}. The results presented in \cite{gevolutiondipole} were additionally obtained by assuming, for simplicity, that $\mathcal{H_O}^3(\mathcal{J_O}-\mathcal{R_O}-1)=0$, with the remark that in the $\Lambda$CDM model, $\hat{\mathcal{J_O}}:=\mathcal{J_O}-\mathcal{R_O}-1\sim 10^{-4}$. As shown in figure \ref{fig:curlJhat}, this combination of parameters fluctuates across the sky with maximum fluctuation amplitude of order $10^2-10^3$ for the N = 64 grid model, i.e. the values found here are up to $10^7$ orders of magnitude above the value of the standard $\Lambda$CDM model. For the N = 8 and N = 4 grids, the fluctuations are reduced to order of 100 and 1-10, respectively. This is still orders of magnitude above the estimate in \cite{gevolutiondipole}. It is therefore not clear that the constraint $\mathcal{H_O}^3\hat{\mathcal{J_O}}=0$ is prudent. However, it must be stressed that even for fixed N, the value of $\hat{\mathcal{J_O}}$ depends strongly on the interpolation scheme. Even for the N = 4 case, the value and even sign of $\mathcal{J_O}$ depends strongly on the interpolation scheme. The value of $\mathcal{R_O}$ is more stable when changing interpolation scheme (see table \ref{table:fit}). For this specific model, the deviation between true and fitted value has nonetheless also been studied, where the fitted value is obtained as $\hat{\mathcal{J_O}}_{\rm fit} = 3\mathcal{Q_O}^2+\mathcal{Q_O}^2-2-6\mathcal{H_O}^2c_3/c $. Already here (for N = 4), there is a significant difference between the value of $\hat{\mathcal{J_O}}$ obtained from fitting and its true value. This is shown in figure \ref{fig:curlJhat_fit}.
\newline\indent
In \cite{ETdipole2, ETdipole1}, none of the approximations listed above were used. Instead, the redshift and distances were computed according to the exact metric of the simulation. The inhomogeneity of the simulation used in \cite{ETdipole1, ETdipole2} is, however, significantly less than in the other models discussed.
\newline\indent

Despite the models studied here and in \cite{gevolutiondipole, ETdipole2, ETdipole1} being quite different, the cosmographic expansion is consistently found to only be valid at very low redshifts, with $z$ of order $0.01$ at most, unless structures are smoothed to very small amplitude/on very large scales.
\newline\newline
In \cite{asha} the general cosmographic expansion was studied using various LTB models made to mimic our cosmic neighborhood. It was found that the radius of convergence was very low (or the rate of convergence very slow) for most of these models and that the second and especially third order cosmographic expansion quickly became very large, leading to the cosmographic expansions deviating by many orders of magnitude from the exact luminosity distance long before $z = 0.1$ was reached. Although the cosmographic expansion is also here found to convergence only at very low redshifts when truncating at third order, the deviations between the cosmographic expansion and the exact luminosity distance are much less significant here than in the most extreme cases found in \cite{asha}. The main reason for this is presumably that the gradients of the relevant spacetime quantities were steeper in the given LTB models than in the models studied here. The fluctuations of $\mathcal{H_O}$ and $\mathcal{Q_O}$ found here are also much smaller than what was found in \cite{asha}.

\section{Summary, discussion and conclusions}\label{sec:summary}
The convergence properties of the general cosmographic expansion introduced in \cite{asta_cosmo} (see also \cite{thesis}) were analyzed in a semi-realistic model of our cosmic neighborhood. This model combines a weak-field relativistic framework with data from CosmicFlows-4. By evaluating the cosmographic expansion along individual lines of sight, it was shown that the expansion can remain non-divergent up to redshifts of $z \sim 0.1$ along certain, special lines of sight. However, along other lines of sight, the expansion diverges at significantly lower redshifts. Except for the smoothest case (N = 4), the results depend strongly on the chosen interpolation scheme. The results also depend quantitatively but not qualitatively on whether grouped or ungrouped CosmicFlows-4 data is used. When examining 768 different lines of sight from a single, realistically placed observer, the mean behavior of the expansion is dominated by those directions where divergence occurs at very low redshifts, because the divergence is so extreme. These results persist when smoothing the studied model by down-sampling the grid, unless a very coarse grid is used.
\newline\indent
Even when N = 8, the cosmographic expansion diverges at low redshifts and the third order cosmographic expansion deviates from the exact redshift-distance relation above 5\% from $z\approx 0.03$ when considering the most extreme lines of sight. However, for an N = 4 grid, corresponding to a smoothing length of 250Mpc/h, the second and third order cosmographic expansions deviate by less than a percent for nearly all light rays up to $z = 0.1$.
\newline\indent
The results obtained here are consistent with previous studies based on more idealized models such as LTB spacetimes or simulations using randomly placed observers, which also find that the cosmographic expansion tends to break down at $z \sim 0.01$ or below. Only when the universe is assumed to be nearly smooth, with fluctuations of order $\delta \sim 0.05$ or on the scale of $125-250$Mpc/h, can the expansion provide accurate approximations at higher redshifts.
\newline\indent
Since only the first three coefficients of the cosmographic expansion were considered, the radius of convergence of the cosmographic expansion was not computed explicitly. It is therefore possible that the breakdown of the expansion is not due to a small radius convergence. Instead, the problem may merely be a slow rate of convergence. In this case, obtaining better agreement between the cosmographic expansion and the exact redshift-distance relation simply requires adding more terms in the expansion.
\newline\indent
The redshift at which the cosmographic expansion ceases to be accurate depends on several factors such as the model universe, its resolution, the observer’s position, and the specific line of sight. Therefore, any practical application of the general cosmographic expansion should be accompanied by explicit convergence tests tailored to the data and model being used. For simulations, such studies are straightforward. For real data, a similar analysis can be performed by constructing a semi-realistic model combining observational input with a weak-field approximation, similar to what was done here.
\newline\newline
Even though the cosmographic expansion itself failed at low redshift in the studied model, the generalized Hubble and deceleration parameters $\mathcal{H_O}$ and $\mathcal{Q_O}$ could be computed directly for each line of sight. Their definitions are not contingent on the convergence of the expansion. In the present analysis, $\mathcal{H_O}$ was found to vary by 10-20\% relative to the background when using the n = 64 grid and linear interpolation, while $\mathcal{Q_O}$ showed variations of up to 4000\% compared to the background deceleration value of $q_0 = -0.55$. These fluctuations reduced by several orders of magnitude when coarsening the grid to N = 4 to obtain a cosmographic expansion that converged in the entire studied interval.
\newline\newline
This work reinforces the conclusion that the general cosmographic expansion of the luminosity distance breaks down at surprisingly low redshifts in realistic cosmic environments. However, it also underscores the substantial information lost when relying solely on the standard FLRW cosmographic framework. In the semi-realistic model studied here, the generalized cosmographic parameters vary across the sky by orders of magnitude, from tens to thousands of percent, when considering the model with the finest grid (N = 64). Such variations are completely masked by the FLRW-based expansion, which averages over these fluctuations. This highlights the importance of understanding how the convergence of the general cosmographic expansion can be improved. Recasting the expansion into a different form or simply including more terms could lead to better performance. Moreover, different observational datasets effectively smooth the universe at different scales. Consequently, some datasets may exhibit better convergence properties. For instance, the Pantheon+ dataset used in conjunction with the general cosmographic expansion in \cite{dipoleH3} may have a larger radius of convergence or a faster rate of convergence than CosmicFlows-4.
\newline\indent
It is important to note that it is currently unclear how the lack of convergence of the cosmographic expansion actually affects constraints of e.g. $\mathcal{H_O}$ and $\mathcal{Q_O}$ obtained using the expansion. Specifically, expansion coefficients obtained from a fit of data to the general cosmographic expansion are not useless even if the cosmographic expansion reproduces the exact redshift-distance relation poorly. The poor convergence rather means that care must be taken when interpreting results, including determining at what scales the coefficients are probing the Universe. When we fit data to a cosmographic expansion, we are in reality ``merely'' fitting the data to a third order polynomial (with vanishing constant term). The fitting coefficients only correspond to those of the cosmographic expansion if the expansion is a good approximation of the redshift-distance relation. However, as pointed out in \cite{gevolutiondipole}, the fitting procedure may be considered as including an implicit smoothing of the data, where the redshift range of the dataset sets the smoothing scale. Thus, the obtained $\mathcal{H_O}$ and $\mathcal{Q_O}$ may be interpreted as representing values in our universe smoothed on some scale. However, as demonstrated here in (table \ref{table:fit}), the fitted coefficients obtained with a fine grid cannot in general be expected to correspond to the actual cosmographic coefficients of a smoothed version of the space. This is important to bear in mind because the cosmographic coefficients have physical meaning and the results obtained here mean that fitted coefficients cannot readily be attributed this same meaning, even in some smoothed sense. Thus, while the idea of \cite{gevolutiondipole} to consider fitted coefficients as representing the expansion of a smoothed spacetime has merit, the exact procedure for smoothing must be established. In relation to this, it may be noted that the smoothing used here was simple down-sampling. Using a more sophisticated smoothing method such as removal of Fourier modes might lead to better agreement between the fitted coefficients of the un-smoothed grid and the true coefficients of the smoothed grid. However, extending a smoothing procedure to construct a coherent, direction-dependent sky map of cosmographic parameters is non-trivial. The challenge becomes even greater in realistic observational settings, where we do not observe continuous $(z, D_L)$ relations along light rays, but rather a discrete and sparse distribution of sources across the sky. In such cases, we aim to extract direction-dependent cosmographic parameters from a set of unconnected data points rather than from smooth curves. Furthermore, the implicit smoothing has implications for handling large datasets; although more data might offer better statistics, it does not guarantee improved precision or accuracy if the fitting procedure inherently smooths the data over large scales. Therefore, careful consideration should be given to how the general cosmographic expansion benefits from increased data volume in the presence of such smoothing.

\begin{acknowledgments}
The author thanks Helene Courtois for correspondence and guidance regarding the density and velocity fields of \cite{fields} and Asta Heinesen, Eoin O Colgain and the anonymous referee for comments on the manuscript. The GNU Scientific Library\footnote{https://www.gnu.org/software/gsl/} was used for solving ordinary differential equations. An array of Python modules were used for plotting and data handling, including NumPy\footnote{https://numpy.org/} \cite{numpy_ref}, Matplotlib\footnote{https://matplotlib.org/} \cite{matplotlib_ref}, healpy\footnote{https://healpy.readthedocs.io/en/latest/} \cite{healpy1, healpy2}, Astropy\footnote{https://www.astropy.org/} \cite{astropy1, astropy2, astropy3}, io\footnote{https://docs.python.org/3/library/os.html},  struct\footnote{https://docs.python.org/3/library/struct.html} and SciPy\footnote{https://scipy.org/citing-scipy/} \cite{scipy_ref}.
\newline\newline
The author is funded by VILLUM FONDEN, grant VIL53032. Part of the presented results were obtained using the UCloud interactive HPC system managed by the eScience Center at the University of Southern Denmark.

\end{acknowledgments}

\appendix

\section{Expressions for generalized Hubble parameter, deceleration parameter, jerk and spatial curvature term}\label{app:curl}
This appendix serves to explicitly show the expressions used for computing the coefficients for the cosmographic expansion of the luminosity distance.
\newline\newline
To compute $\mathcal{H}$ we need expressions for the expansion rate, shear and acceleration. The expansion rate is, for the considered weak-field model, given by
\begin{align}
    \Theta = \nabla_\alpha u^\alpha\approx 3H + \partial_i u^i,
\end{align}
where $u^\alpha \approx (1, v^i)$ at first order (with $v^i$ the peculiar velocities obtained from the CosmicFlows-4 data). The non-vanishing components of the shear are
\begin{align}
\begin{split}
\sigma_{ii} &= \frac{2}{3}a^2\partial_i v^i\\
\sigma_{ij, (i\neq j)} & = \frac{1}{2}a^2\left(\partial_i v^j + \partial_j v^i\right),
\end{split}
\end{align}
where sums over repeated indexes are {\em not} implied. The 4-acceleration has components given by
\begin{align}
\begin{split}
    a_t & = -aa_{,t}\sum (v^i)^2\\
    a_i &= 2aa_{,t}v^i + a^2(\partial_t v^i + v^j\partial_j v^i),
\end{split}
\end{align}
where sums {\em are} implied over repeated indexes and where $v^i_{,t} = 0$ was assumed during computations.
\newline\newline
With the above we can compute $\mathcal{H}, \mathcal{Q}$ and $\mathcal{J}$. To compute $\mathcal{R}$ we further need
\begin{align}
\begin{split}
        k^\mu k^\nu R_{\mu\nu} &= \frac{8\pi G}{c^4}\rho k^\mu k^\nu \left(u_{\mu}u_{\nu}+\frac{1}{2}g_{\mu\nu}\right)\\
        & = \frac{8\pi G\rho}{c^4}\left(u^\mu k_\nu\right)^2,
\end{split}
\end{align}
where the first equality comes from employing Einstein's equation.

\section{Density field}\label{app:density}
\begin{figure*}
    \centering
    \includegraphics[width=\columnwidth]{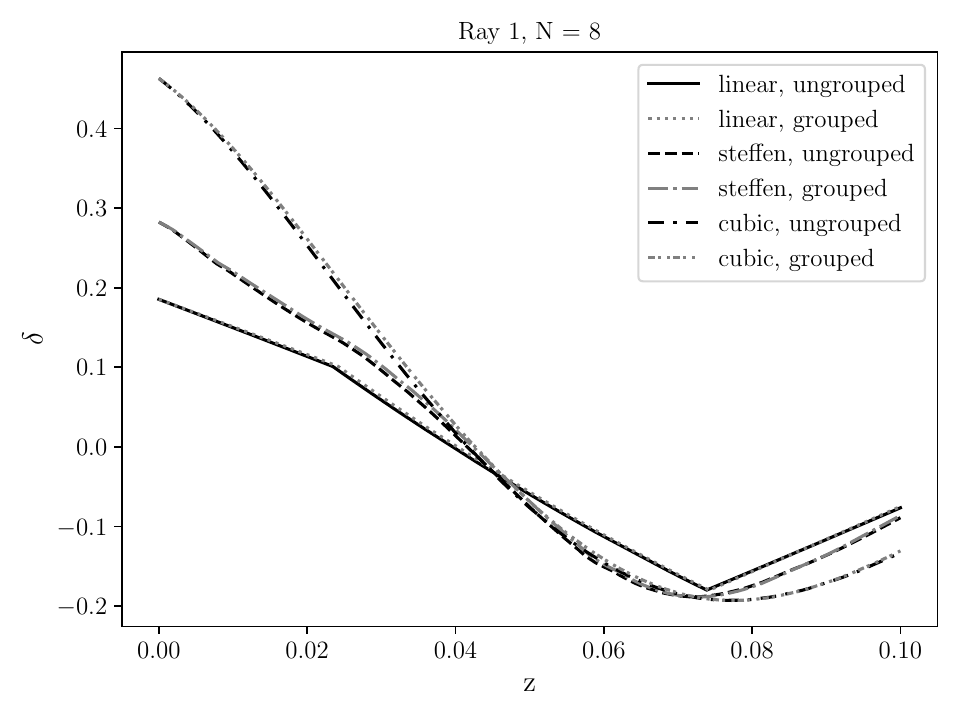} 
    \includegraphics[width=\columnwidth]{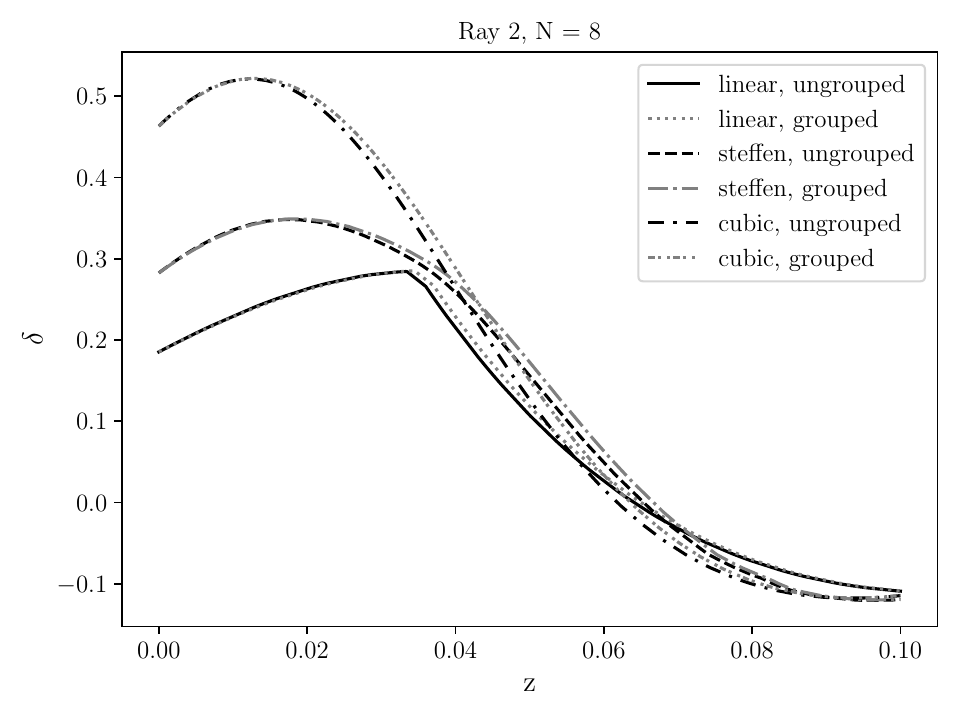}\\
    \includegraphics[width=\columnwidth]{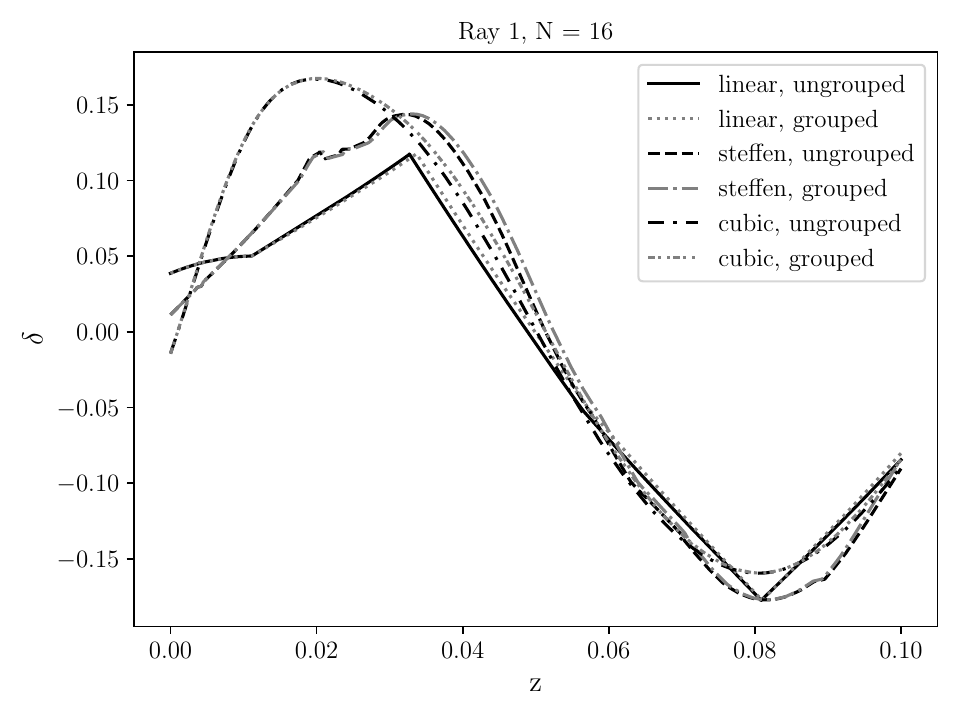}
    \includegraphics[width=\columnwidth]{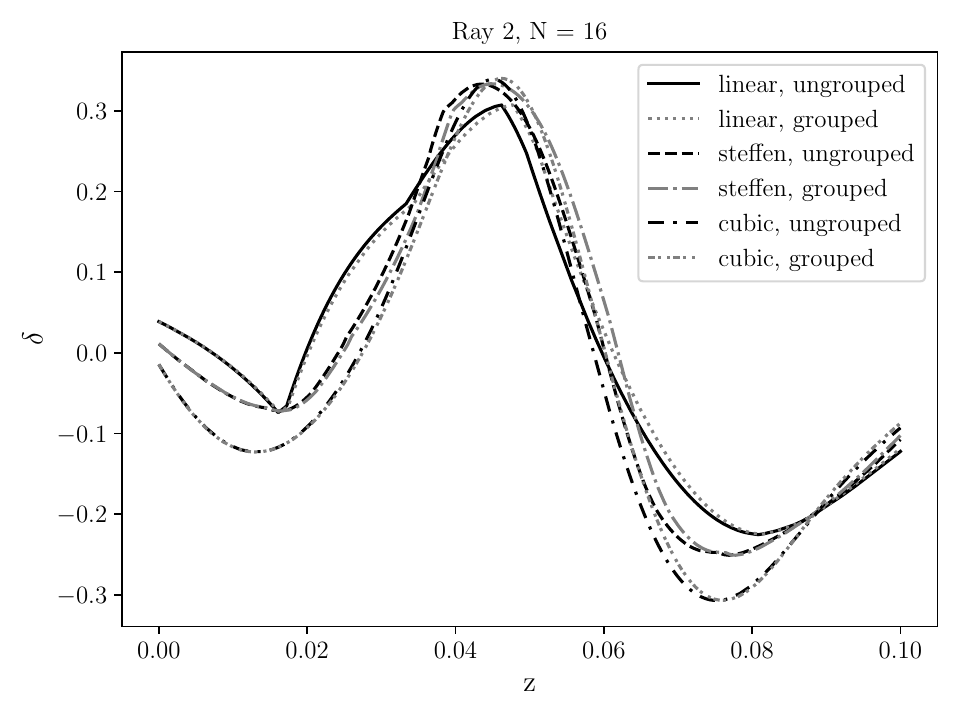}\\
    \caption{The density contrasts along ray 1 and 2 using N = 8 and 16 grids with different interpolation schemes.}
    \label{fig:ray1ray2_delta}
\end{figure*}

For the interested reader, the density fields along the two fiducial light rays (ray 1 and ray 2) are shown in figure \ref{fig:ray1ray2_delta} for the N = 8 grid using three different interpolation schemes (linear, cubic and Steffen).

\clearpage

\begin{thebibliography}{99}
\bibitem{dipoleH1} Mohamed Rameez, Subir Sarkar, Is there really a Hubble tension?, Class. Quantum Grav. 38 (2021) 154005, arXiv:1911.06456v4 [astro-ph.CO]
\bibitem{dipoleH2} Francesco Sorrenti, Ruth Durrer, Martin Kunz, The Dipole of the Pantheon+SH0ES Data,  JCAP11(2023)054, arXiv:2212.10328v2 [astro-ph.CO]
\bibitem{dipoleH3} Jessica A. Cowell, Suhail Dhawan, Hayley J. Macpherson, Potential signature of a quadrupolar Hubble expansion in Pantheon+ supernovae, MNRAS 526, 1482–1494 (2023), arXiv:2212.13569v1 [astro-ph.CO]
\bibitem{dipoleH4} Leandros Perivolaropoulos, On the isotropy of SnIa absolute magnitudes in the Pantheon+ and SH0ES samples, Phys. Rev. D 108, 063509 (2023), arXiv:2305.12819v1 [astro-ph.CO]
\bibitem{dipoleH5} Li Tang, Hai-Nan Lin, Liang Liu, Xin Li, Consistency of Pantheon+ supernovae with a large-scale isotropic universe, Chinese Phys. C 47 125101,  arXiv:2309.11320v2 [astro-ph.CO]
\bibitem{dipoleH6} Ruairi McConville, Eoin O Colgain, Anisotropic Distance Ladder in Pantheon+ Supernovae, Phys. Rev. D 108, 123533 (2023), arXiv:2304.02718v2 [astro-ph.CO]
\bibitem{dipoleH7} Jianping Hu, Jian Hu, Xuandong Jia, Baoquan Gao, Fayin Wang, Testing cosmic anisotropy with Pade approximation and Pantheon+ sample, A\&A 689, A215 (2024), arXiv:2406.14827v1 [astro-ph.CO] 
\bibitem{dipoleH8} Carlos A. P. Bengaly, Cassio Pigozzo, Jailson S. Alcaniz, Testing the isotropy of cosmic acceleration with Pantheon+ and SH0ES: A cosmographic analysis, Phys.Rev.D 109 (2024) 12, 123533, arXiv:2402.17741v2 [astro-ph.CO] 
\bibitem{dipoleH9} J. P. Hu, Y. Y. Wang, J. Hu, F. Y. Wang, Testing the cosmological principle with the Pantheon+ sample and the region-fitting method, A\&A, 681, A88 (2024), arXiv:2310.11727v2 [astro-ph.CO]
\bibitem{dipoleH_summary} Animesh Sah, Mohamed Rameez, Subir Sarkar, Christos Tsagas, Anisotropy in Pantheon+ supernovae, arXiv:2411.10838v1 [astro-ph.CO]
\bibitem{review} Mohamed Rameez, Anisotropy in the cosmic acceleration inferred from supernovae, arXiv:2411.14758v1 [astro-ph.CO]

\bibitem{ani1} Nathan Secrest, Sebastian von Hausegger, Mohamed Rameez, Roya Mohayaee, Subir Sarkar, A Challenge to the Standard Cosmological Model, Astrophys. J. Lett. 937 (2022) L31, arXiv:2206.05624v2 [astro-ph.CO]
\bibitem{ani2} Lawrence Dam, Geraint F. Lewis, Brendon J. Brewer, Testing the Cosmological Principle with CatWISE Quasars: A Bayesian Analysis of the Number-Count Dipole, MNRAS 525, 231–245 (2023), arXiv:2212.07733v2 [astro-ph.CO]
\bibitem{ani3} J. D. Wagenveld, H-R. Klöckner, D. J. Schwarz, The cosmic radio dipole: Bayesian estimators on new and old radio surveys, A\&A 675, A72 (2023), arXiv:2305.15335v1 [astro-ph.CO]
\bibitem{ani_original} G. F. R. Ellis, J. E. Baldwin, On the expected anisotropy of radio source counts, MNRAS (1984) 206, 377-381, https://doi.org/10.1093/mnras/206.2.377

\bibitem{bulkflow} Richard Watkins et al., Analyzing the Large-Scale Bulk Flow using CosmicFlows4: Increasing Tension with the Standard Cosmological Model, MNRAS 524, 1885–1892 (2023), arXiv:2302.02028v1 [astro-ph.CO]


\bibitem{asta_cosmo} Asta Heinesen, Multipole decomposition of the general luminosity distance 'Hubble law' -- a new framework for observational cosmology, JCAP05(2021)008, arXiv:2010.06534v2 [astro-ph.CO]
\bibitem{thesis} Umeh, O. 2013. The influence of structure formation on the evolution of the universe. University of Cape Town (PhD thesis)

\bibitem{augment1} Roy Maartens, Jessica Santiago, Chris Clarkson, Basheer Kalbouneh, Christian Marinoni, Covariant cosmography: the observer-dependence of the Hubble parameter, JCAP09(2024)070, arXiv:2312.09875v3 [astro-ph.CO]
\bibitem{augment2} Basheer Kalbouneh, Christian Marinoni, Roy Maartens, Cosmography of the Local Universe by Multipole Analysis of the Expansion Rate Fluctuation Field, arXiv:2401.12291v1 [astro-ph.CO]
\bibitem{augment3} Basheer Kalbouneh, Jessica Santiago, Christian Marinoni, Roy Maartens, Chris Clarkson, Maharshi Sarma, Expanding covariant cosmography of the local Universe: incorporating the snap and axial symmetry, arXiv:2408.04333v1 [astro-ph.CO]

\bibitem{early1} Chris Clarkson, Obinna Umeh, Is backreaction really small within concordance cosmology?, Class. Quantum Grav. 28 164010 (2011), arXiv:1105.1886v1 [astro-ph.CO]
\bibitem{early2} Stella Seitz, Peter Schneider, Juergen Ehlers, Light Propagation in Arbitrary Spacetimes and the Gravitational Lens Approximation, Class.Quant.Grav.11:2345-2374,1994, arXiv:astro-ph/9403056v1
\bibitem{early3} G.F.R. Ellis, S.D. Nel, R. Maartens, W.R. Stoeger, A.P. Whitman, Ideal observational cosmology, Physics Reports, Volume 124, Issues 5–6, July 1985, Pages 315-417

\bibitem{ETdipole1} Hayley J. Macpherson, Asta Heinesen, Luminosity distance and anisotropic sky-sampling at low redshifts: a numerical relativity study, Phys. Rev. D 104, 023525 (2021), arXiv:2103.11918v3 [astro-ph.CO]
\bibitem{ETdipole2} Hayley J. Macpherson, Cosmological distances with general-relativistic ray tracing: framework and comparison to cosmographic predictions, JCAP03(2023)019, arXiv:2209.06775v2 [astro-ph.CO]

\bibitem{gevolutiondipole} Julian Adamek, Chris Clarkson, Ruth Durrer, Asta Heinesen, Martin Kunz, Hayley J. Macpherson, Towards Cosmography of the Local Universe, 	The Open Journal of Astrophysics 7 (2024), arXiv:2402.12165v2 [astro-ph.CO]



\bibitem{cosmoFLRW1} Celine Cattoen, Matt Visser, The Hubble series: Convergence properties and redshift variables, Class.Quant.Grav.24:5985-5998,2007, arXiv:0710.1887v1 [gr-qc]

\bibitem{cosmoFLRW2} E. O Colgain, M.M. Sheikh-Jabbari, Elucidating cosmological model dependence with $H_0$, Eur.Phys.J.C 81 (2021) 10, 892, arXiv:2101.08565v3 [astro-ph.CO]

\bibitem{asha} Asha B. Modan, S. M. Koksbang, On the convergence of cosmographic expansions in Lemaitre-Tolman-Bondi models, Class. Quantum Grav. 41 235018 (2024), arXiv:2408.07459v2 [gr-qc]



\bibitem{LTB1} G. Lemaitre: L'Universe en expansion, Annales de la Societe Scientifique de Bruxelles A 53, 51 (1933), English translation: The expanding universe, Gen. Rel. Grav. 29, 637 (1997)
\bibitem{LTB2} R. C. Tolman: Effect of Inhomogeneity on Cosmological Models, Proc. Natl. Acad. Sci. USA
20, 169-176 (1934)
\bibitem{LTB3} H. Bondi: Spherically Symmetrical Models in General Relativity, Month. Not. Roy. Astr. Soc. 107,410 (1947)

\bibitem{fields} H. M. Courtois, A. Dupuy, D. Guinet, G. Baulieu, F. Ruppin, P. Brenas, Gravity in the Local Universe : density and velocity fields using CosmicFlows-4, A\&A 670, L15 (2023), arXiv:2211.16390v4 [astro-ph.CO]
\bibitem{CosmicFlows4} R. Brent Tully et al., Cosmicflows-4, 2023 ApJ 944 94, arXiv:2209.11238v2 [astro-ph.CO]
\bibitem{basin} Aurelien Valade, Noam I. Libeskind, Daniel Pomarede, R. Brent Tully, Yehuda Hoffmann, Simon Pfeifer, Ehsan Kourkchi, Identification of Basins of Attraction in the Local Universe, Nat. Astron. (2024), arXiv:2409.17261v1 [astro-ph.CO]
\bibitem{group} R. Graziani, H. M. Courtois, G. Lavaux, Y. Hoffman, R. B. Tully, Y. Copin, D. Pomarede, The peculiar velocity field up to z$\sim$0.05 by forward-modeling Cosmicflows-3 data, MNRAS 488, 5438–5451 (2019), arXiv:1901.01818v1 [astro-ph.CO] 

\bibitem{gevolution} Julian Adamek, David Daverio, Ruth Durrer, Martin Kunz, gevolution: a cosmological N-body code based on General Relativity, JCAP 1607 (2016) no.07, 053, arXiv:1604.06065v2 [astro-ph.CO]


\bibitem{greenwald} S. R. Green and R. M. Wald, Newtonian and Relativistic Cosmologies, Phys. Rev. D85 (2012) 063512, arXvi:1111.2997

\bibitem{mapLTB1} S. M. Koksbang, S. Hannestad, Methods for studying the accuracy of light propagation in N-body simulations, Phys. Rev. D 91, 043508 (2015), arXiv:1501.01413v3 [astro-ph.CO]
\bibitem{mapLTB2} S. M. Koksbang, S. Hannestad, Studying the precision of ray tracing techniques with Szekeres models, Phys. Rev. D 92, 023532 (2015), arXiv:1506.09127v3 [astro-ph.CO]

\bibitem{god_bog} George F. R. Ellis, Roy Maartens and Malcolm A. H. MacCallum, Relativistic Cosmology, Cambdirge University Press, ISBN 978-0-521-38115-4, 2012


\bibitem{bright_side} Krzysztof Bolejko, Chris Clarkson, Roy Maartens, David Bacon, Nikolai Meures, Emma Beynon, Anti-lensing: the bright side of voids, 	Phys. Rev. Lett. 110, 021302 (2013), arXiv:1209.3142v3 [astro-ph.CO]

\bibitem{bonvin} Camille Bonvin, Effect of Peculiar Motion in Weak Lensing, Phys.Rev.D78:123530,2008, arXiv:0810.0180v3 [astro-ph]

\bibitem{sachs} R. Sachs,  Gravitational Waves in General Relativity. VI. The Outgoing Radiation Condition, Proceedings of the Royal Society of London Series A, 264(1318):309–338 (1961)



\bibitem{pade_summary} Hao Wei, Xiao-Peng Yan, Ya-Nan Zhou, Cosmological Applications of Pade Approximant, JCAP1401:045,2014, arXiv:1312.1117v3 [astro-ph.CO]

\bibitem{resolution} Louis Coates, Julian Adamek, Philip Bull, Caroline Guandalin, Chris Clarkson, Observing relativistic features in large-scale structure surveys -- II: Doppler magnification in an ensemble of relativistic simulations, MNRAS,504(3),2021,3534-3543, arXiv:2011.12936v2 [astro-ph.CO]


 \bibitem{numpy_ref} Harris, C.R., Millman, K.J., van der Walt, S.J. et al., Array programming with NumPy, Nature 585, 357–362 (2020), arXiv:2006.10256v1 [cs.MS]

\bibitem{matplotlib_ref} J. D. Hunter, Matplotlib: A 2D Graphics Environment, Computing in Science \& Engineering, vol. 9, no. 3, pp. 90-95, 2007.

\bibitem{healpy1} K. M. Gorski, E. Hivon, A. J. Banday, B. D. Wandelt, F. K. Hansen, M. Reinecke, M. Bartelman, Astrophys.J.622:759-771,2005, arXiv:astro-ph/0409513v1

\bibitem{healpy2} Andrea Zonca, Leo Singer, Daniel Lenz, Martin Reinecke, Cyrille Rosset, Eric Hivon, Krzysztof Gorski, healpy: equal area pixelization and spherical harmonics transforms for data on the sphere in Python, Journal of Open Source Software, vol. 4, issue 35, id. 1298 (2019)


\bibitem{astropy1} Thomas P. Robitaille et al., Astropy: A community Python package for astronomy, A\&A 558, A33 (2013), arXiv:1307.6212v1 [astro-ph.IM]

\bibitem{astropy2} The Astropy Collaboration, The Astropy Project: Building an inclusive, open-science project and status of the v2.0 core package, 2018 AJ 156 123, arXiv:1801.02634v2 [astro-ph.IM] 

\bibitem{astropy3} The Astropy Collaboration, The Astropy Project: Sustaining and Growing a Community-oriented Open-source Project and the Latest Major Release (v5.0) of the Core Package, ApJ 935 167, arXiv:2206.14220v1 [astro-ph.IM]

\bibitem{scipy_ref} Pauli Virtanen et al., SciPy 1.0: Fundamental Algorithms for Scientific Computing in Python. Nature Methods, 17(3), 261-272 (2020), arXiv:1907.10121v1 [cs.MS]










\end{thebibliography}
\end{document}